\documentclass[pre,twocolumn,preprintnumbers,amsmath,amssymb,floatfix]{revtex4}

\usepackage{subfigure}

\usepackage{graphicx}
\usepackage{dcolumn}
\usepackage{bm}
\usepackage{hyperref}
\usepackage{enumerate}
\usepackage{xcolor}

\def\be{\begin{equation}}
\def\ee{\end{equation}}

\def\ovm#1{\bgroup \color{blue} OVM: #1\egroup}

\let\theta\vartheta

\newcommand{\mcm}[1]{{\color{black} #1}}
\newcommand{\mcmch}[1]{{\color{black} #1}}

\newcommand{\hh }[1]{ \hat{\bm{#1}} }
\newcommand{\m }[1]{ \mathbf{#1} }

\newcommand{\rr}{{\mathbf r}}

\newcommand{\w}{\mbox{\boldmath$\omega$}}

\definecolor{darkgreen}{rgb}{0,0.5,0}

\newcommand{\matteop}[1]{{\color{black} #1}}
\newcommand{\cris}[1]{{\color{black} #1}}
\newcommand{\marco}[1]{{\color{black} #1}}

\newcommand{\BEQ}{\begin{equation}}
\newcommand{\EEQ}{\end{equation}}
\newcommand{\BEA}{\begin{eqnarray}}
\newcommand{\EEA}{\end{eqnarray}}

\def\be{\begin{equation}}   \def\ee{\end{equation}}
\def\eq#1{{Eq.(\ref{#1})}}

\newcommand{\GG}{\textcolor{black}{G}}
\newcommand{\RRR}{\textcolor{black}{R}}

\begin{document}

\title{Information and motility exchange in collectives of active particles}

\author{ Matteo Paoluzzi$^{1,2}$}
\email{mttpaoluzzi@gmail.com}

\author{ Marco Leoni$^{3}$}
\email{leoni@lal.in2p3.fr}

\author{ M. Cristina Marchetti$^{4}$ }

\affiliation{
$^1$ ISC-CNR,  Institute  for  Complex  Systems,  Piazzale  A.  Moro  2,  I-00185  Rome,  Italy \\
$^2$ Dipartimento di Fisica, Sapienza University of Rome, Piazzale  A.  Moro  2, I-00185, Rome, Italy \\
$^3$ Universit\'e  Paris-Saclay, CNRS, Laboratoire de l'acc\'el\'erateur lin\'eaire, 91898, Orsay,  France \\
$^4$ Department of Physics, University of California Santa Barbara, Santa Barbara, CA 93106, USA
}

\date{\today}

\begin{abstract}
We examine the interplay of motility and information exchange  in a model of run-and-tumble active particles  where the particle's motility is encoded as a bit of information 
 that can be exchanged upon contact according to the rules of AND and OR logic gates in a circuit.
Motile AND  particles become 
 non-motile upon contact with a non-motile particle.  Conversely, motile OR particles remain motile upon collision with their non-motile counterparts.
AND particles that have become non-motile  additionally \mcm{``reawaken'',  i.e.,} recover their motility,  at a fixed rate $\mu$, as in
the SIS
(Susceptible, Infected, Susceptible) model of epidemic spreading, where an  infected agent 
  can become healthy again, but keeps no memory of the recent infection, hence it is susceptible to
a renewed infection. For $\mu=0$,  both AND and OR particles relax irreversibly to absorbing states of all non-motile or all motile particles, respectively. \mcm{T}he relaxation kinetics  
is\mcm{, however,} faster for OR particles that remain active throughout the process. At finite $\mu$, the AND dynamics is controlled by the interplay between reawakening and collision rates. The system evolves to a state of all motile particles (an absorbing state in the language of absorbing phase transitions)  for $\mu>\mu_c$ 
and to a mixed state with coexisting motile and non-motile particles (an active state in the language of absorbing phase transitions) for $\mu<\mu_c$. 
The final state exhibits a rich structure controlled by motility-induced aggregation. Our work can be relevant to biochemical signaling in motile bacteria, the spreading of epidemics and \mcm{of} social consensus, as well as light-controlled organization of active colloids.  
\end{abstract}

\maketitle

\section{Introduction}
 \begin{figure}[!h]
\includegraphics[width=.9\columnwidth]{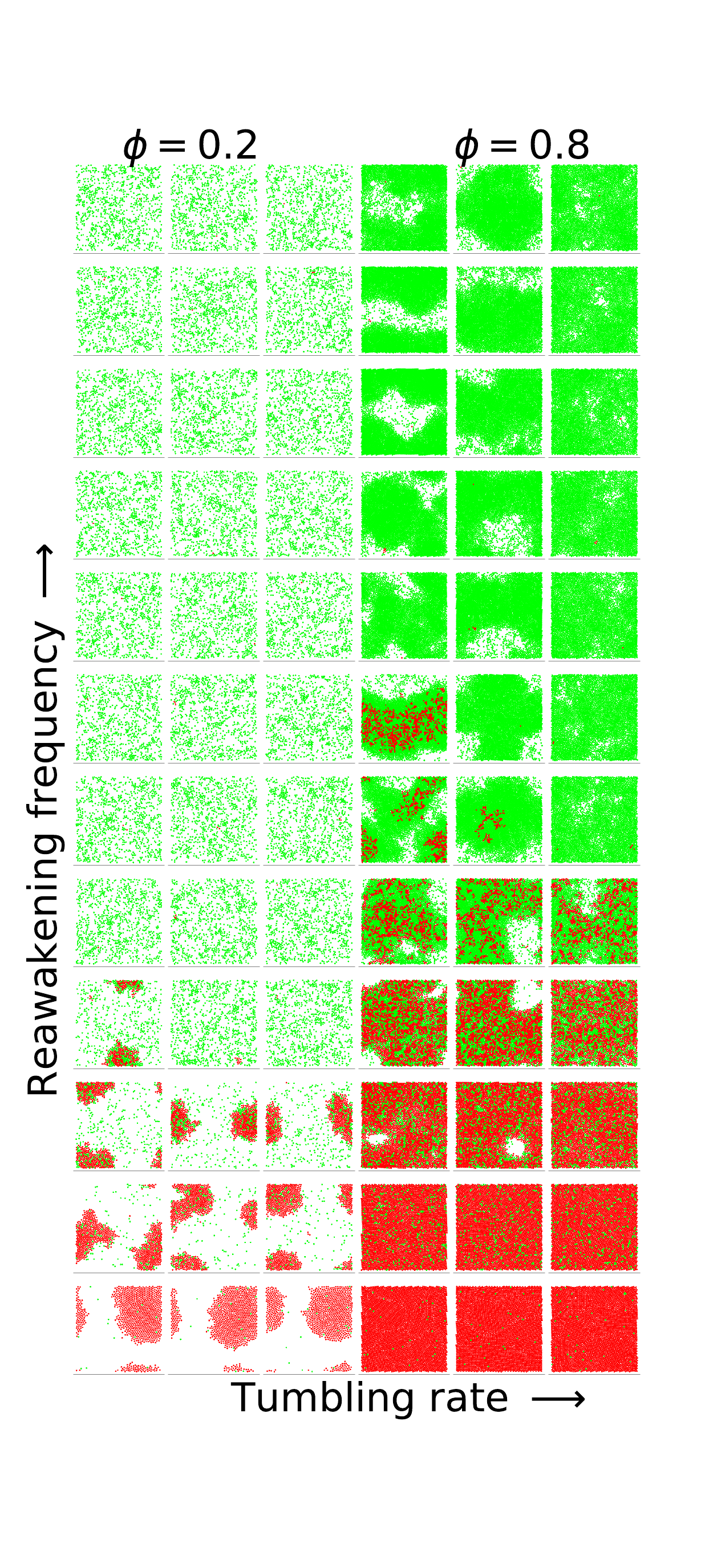} 
\caption{ Representative snapshots  of the final configuration of AND particles for two values of the total packing fraction $\phi$, $\phi=0.2$ (left) and $\phi=0.8$ (right). 
Red particles are nonmotile, green particles
are motile. The tumbling rate is $\lambda=0.01,0.1,1$ from left to right at each packing fraction. At both packing fractions the system escapes the absorbing state as the awakening rate is increased. 
At low packing fraction, the mixed state  separates into a cluster \mcm{of} non-motile particles surrounded by a gas of motile ones. At high packing fraction, motile particles in the active state exhibit MIPS, while the mixed state is homogeneous. 
\label{fig:snap} }
\end{figure}

Swimming  bacteria and living  cells  are examples of entities that consume energy to generate  autonomous motion.
Through interactions, these systems organize in complex 
 patterns on scales much larger than those of the individual constituents.
 This behavior has provided inspiration for the development of the field of    
 active matter that has had remarkable success at describing some of the spontaneous organization seen in nature on many scales, from the flocking of birds 
 to the collective migration of epithelial cells in wound healing \cite{Marchetti-Rev-Mod-Phys,cates_rep,Ramaswamy,RevModPhys.88.045006}.  
 
 So far most active matter studies have focused on the  role of reciprocal mechanical or 
 rule-based interactions, such as steric repulsion or medium-mediated hydrodynamic couplings, 
 in controlling the emergence of nonequilibrium collective behavior. 
 But living entities often interact through the exchange of information transmitted through biochemical 
 signaling, chemical, visual and other clues. Information exchange carried by motile individuals is often 
 non-reciprocal and its role in controlling emergent structures in collections of motile agents is only beginning to be explored \cite{smeets2016emergent,PhysRevResearch.1.023026,PhysRevX.7.011028,PhysRevE.97.042604,sibona2012influence,PhysRevLett.100.168103,peruani2019reaction}. 
 
 \mcm{Examples} of information exchange common in nature \mcm{are} biochemical signaling, that controls communication among microbes~ \cite{keller2006communication,hibbing2010bacterial}, and chemotaxis~\cite{wadhams2004making,van2004chemotaxis}, \mcm{that drives} the formation of  complex patterns \cite{PhysRevLett.123.018101}.
\mcm{Both have} been modeled extensively
by coupling a diffusive agent to continuum models of  active gels~\cite{Grill} and  
agent-based simulation~\cite{Libchen}. 
Closer in spirit \mcm{to} the  model considered here is
 recent experimental  work
on active colloids where activity \mcm{is} controlled by an external 
feedback loop that tunes the light intensity responsible for driving the colloids'  motility  \cite{Solano, Bauerle}, mimicking the use of light intensity  to control and trigger collective behavior \mcm{of} photokinetic bacteria \cite{frangipane2018dynamic,arlt2018painting}.
More sophisticated examples include light-activated colloidal particles  able to {\it learn} and to share information with other particles
while interacting with the external environment \cite{Learning}. 
Finally, information exchange among motile individuals is directly relevant to the understanding of epidemic spreading, social dynamics and robotic communication \cite{rubenstein2014programmable,PhysRevE.97.042604,agliari2006efficiency,ferguson2007capturing,RevModPhys.87.925}. 

Much quantitative understanding of the behavior of active systems has come \mcm{from}  minimal models of Active Brownian or Run-and-Tumble Particles (ABP or RTP) consisting of collections of self-propelled spherical particles with purely repulsive interactions that propel themselves at fixed speed and change direction through rotational noise or tumbling events.  
Building on this body of work, we recently considered a minimal model of active agents where the particle's motility is encoded as a bit of information 
 that can be exchanged upon contact interaction
according to logic rules corresponding to AND and OR gates in an electronic circuit~\cite{Draft-Matteo}.
Motile particles obeying AND rules always lose their motility upon interaction with non-motile ones. In this case
an initial state of one non-motile particle in a sea of motile ones  always evolves to an absorbing state where all particles are nonmotile. 
Conversely, OR motile particles remain motile when interacting with nonmotile ones. In this case an initial state of one motile particle in a sea of non-motile ones  always evolves towards 
the absorbing state where all particles are motile. 
This model is analogous to SI models (S, susceptible, I, infected) 
of epidemic spreading where information is transmitted only in one direction and irreversibly, 
with motile particles corresponding to healthy agents and non-motile particles to infected ones.

 In the present paper we examine a richer model analogue to SIS (Susceptible, Infected, Susceptible) models of epidemic spreading \cite{Murray}, where an  infected agent 
  can become healthy again, but keeps no memory of the recent infection, hence it is susceptible to
 a renewed infection~ \cite{PhysRevLett.100.168103}. 
   We do this by allowing non-motile particles to regain their motility or ``re-awaken'' at an average rate $\mu$.
   The dynamics is then controlled by the interplay between reawakening rate and collision rate, with a critical value $\mu_c$ of reawakening rate controlling the properties of the final steady state.
  For $\mu \geq \mu_c$ the system reaches an absorbing state where all particles are motile. 
   In the terminology of absorbing state phase transitions,  this state, although composed entirely of motile particles, is {\it inactive} because the spreading of nonmotile particles has ceased
  ~ \cite{RevModPhys.76.663,hinrichsen2000non}. 
   For $0<\mu < \mu_c$ we have a mixed state of motile and nonmotile particles. Again, in the language of absorbing states, this is an {\it active} state because 
   it is a dynamical steady state where particles continue to exchange their motility.  
   To avoid confusion, we will refer to these two states as \emph{motile} and \emph{mixed}, respectively.
  For $\mu=0$ the system evolves at long times to the absorbing state where all particles are non-motile. Our simulation suggests that this sate only exists for $\mu=0$, and  that the system remains mixed for any finite value of $\mu<\mu_c$. This could, however, be a consequence of the unavoidable finite time scale of our simulations. Whether there is a lower, but finite critical value \matteop{$\mu_c$} below which the system reaches the absorbing non-motile state remains an open question.
  \mcm{We note that previous work on absorbing states in active systems has focused on particles with infinite run length, where the system can get trapped in active absorbing states~\cite{reichhardt2014absorbing}.}

\begin{figure}[!tb]
\includegraphics[width=1.\columnwidth]{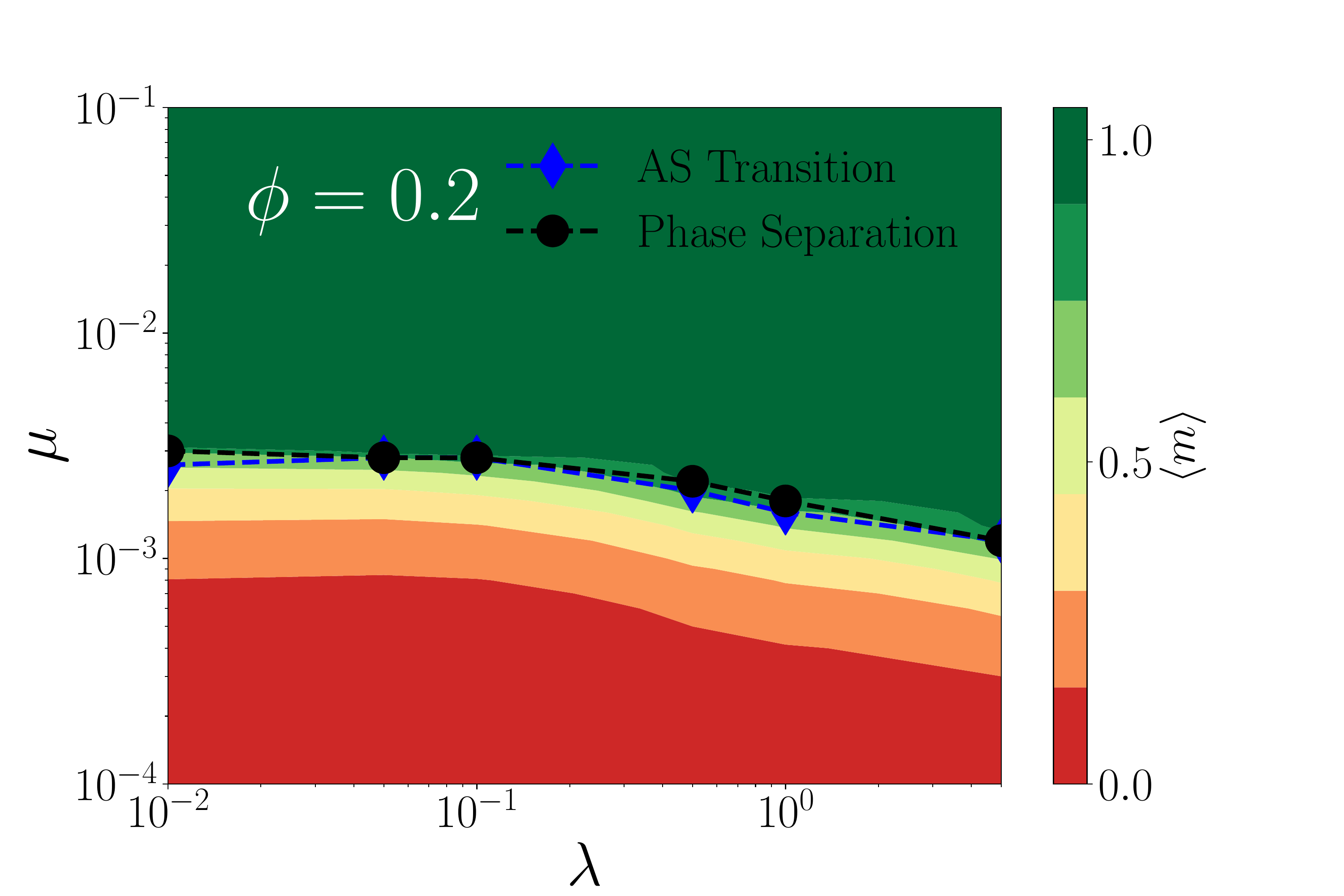}\\
\includegraphics[width=1.\columnwidth]{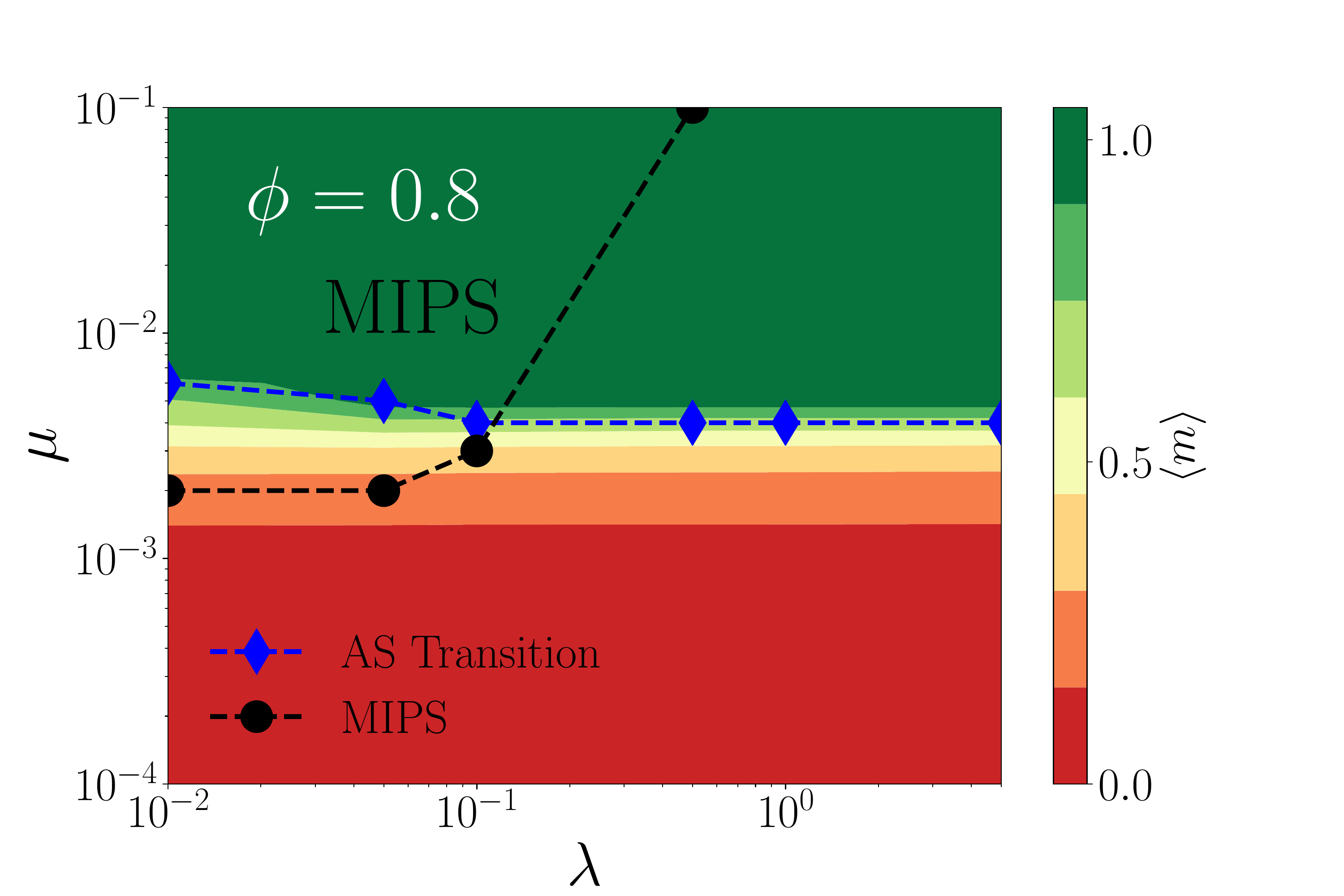} 
\caption{ Phase diagram for \mcm{packing fraction} $\phi=0.2$ (top) and $\phi=0.8$ (bottom) in the \mcm{plane of tumbling rate $\lambda$ and re-awakening rate $\mu$.} The blue diamonds correspond to the \mcm{Absorbing State (AS) } transition between mixed and active states identified \mcm{by} the location of the peak in the 
susceptibility $\chi_s$. The black circles are the points where the Binder cumulant $g_4(\lambda,\mu)$ \cite{binder1981finite,rovere1988block,rovere1993computer} becomes
negative, signaling the onset of  \matteop{MIPS for $\phi=0.8$ and \mcm{of }{ \color{black}{de-mixing of}}  motile and non-motile particles for $\phi=0.2$.}
\label{fig:psi02} }
\end{figure}

A pictorial phase diagram  depicting snapshots of  representative steady-state configurations is shown in Fig.~\ref{fig:snap}.
Both motile and mixed states show a rich spatial organization. In the motile state at $\mu \geq \mu_c$, 
motile particles exhibit motility-induced phase separation (MIPS)~\cite{Tailleur08,Fily2012,cates2015motility} at high density and low tumbling rates.  
In the mixed state, we find spatial patterns at both low and high total density.
 At low densities  nonmotile particles form a cluster surrounded by a gas of motile particles.  
At high densities the interplay between aggregation of non-motile particles and MIPS results in the opening of bubbles 
in a mixed background resembl\mcm{ing} a reverse MIPS, as observed in single components ABPs at very high density and motility~\cite{Fily2014,cates2015motility,patch2018curvature}. 
This effect is most pronounced at low tumbling rates, where the dynamics is most persistent, suggesting that it is indeed driven by motility.
\mcm{A quantitative phase diagram depicting the various regimes as function of the re-awakening and tumbling rates is shown in Fig.~\ref{fig:psi02}.}
%

%
Our  work \mcm{demonstrates} that the interplay of motility and information spreading is responsible for complex spatial structures that can be controlled by tuning the 
total particle density and the re-awakeining rate. \mcm{Our} model may \mcm{therefore} be relevant to recent experimental work on engineered active particles where the motility of individual particles can be controlled optically. 
It also provides a new approach to problems such as epidemics  or opinion spreading \cite{Keeling,opinion,RevModPhys.87.925}. 
While these problems have been studied extensively, most previous work has been carried out for 
agents sitting on a network of fixed connectivity \cite{PhysRevLett.109.128702,riley2007large}.
In our model, in contrast, the connectivity changes in time as it is determined by the agent dynamics, and the properties of such dynamics affect information spreading. 
Finally, the model could be adapted to describe the exchange of other internal traits other than motility.
In certain  bacteria or eukaryotic cells contact interactions are in fact needed for the exchange of chemical signaling, as is the case for instance
for C-signaling that mediates collective motility in the bacteria  \emph{Myxococcus Xantus}.

The details of the agent-based model are introduced in Section II. %
In Section III we present the results of the numerical simulations and
the metrics used to construct quantitative phase diagrams and to characterize the relaxation kinetics and the structure of the steady states. 
In Section IV we discuss a continuum model  
and conclude with a few remarks in Section V.

  \section{Model}
  \label{Sec:model} 
We consider $N$ spherical particles of diameter $a$ performing run-and-tumble dynamics in two dimensions.
The particles have identical size, self propulsion \mcm{speed}, and tumbling rate. \mcm{T}hey interact via short-range repulsive interactions.
They are only distinguished by their motility state (motile or non-motile) represented by their color (green \mcm{(G)} for motile and red \mcm{(R)} for non-motile). 
Particles exchange their motility state upon collision according to rules inspired by logic gates. 
The motility state of AND particles evolves according to the rules associated with the AND gate of a logic
 circuit that gives a high output only if all its inputs are high.
Conversely, OR particles evolve according to the rules associated with an OR gate that gives a high output if one or more of its inputs are high.
The corresponding interaction rules are given in Table \ref{table1}. 

\begin{table}
\centering
      \begin{tabular}{|c|c|}
      \hline
   AND & OR     \\  \hline
  $\GG+\RRR \rightarrow \RRR+\RRR$ & $\GG+\RRR \rightarrow \GG+\GG$\\
  $\RRR+\GG \rightarrow \RRR+\RRR$ & $\RRR+\GG \rightarrow \GG+\GG$\\
  $\GG+\GG \rightarrow \GG+\GG$ & $\GG+\GG \rightarrow \GG+\GG$\\
  $\RRR+\RRR \rightarrow \RRR+\RRR$ & $\RRR+\RRR  \rightarrow \RRR+\RRR$\\
  \hline
   \end{tabular}
      \caption{The logic interaction rules that control motility exchange among  non-motile/red ($\RRR$) and motile/green  ($\GG$) particles interacting via AND and OR rules. }
         \label{table1}
 \end{table}
  %

In the case of AND rules, non-motile (\mcm{R}) particles can also re-acquire their motility, or ``reawaken'', at a rate $\mu$. When $\mu=0$ the loss/gain of motility is analogue to the spreading of disease in the Susceptible-Infected (SI) model of epidemics dynamics, with motile particles corresponding to infected individuals and \textcolor{black}{nonmotile particles} to susceptible individuals. For finite values of $\mu$ the model corresponds to the Susceptible-Infected-Susceptible (SIS) model of epidemics spreading \cite{Murray}. 

\paragraph*{Agent-based model.} Denoting the motility state of particle $i$ by an internal variable $\sigma_i(t)$, with 
 $\sigma = {0, 1}$ for non-motile and motile particles, respectively, the dynamics of the system is described by the equations
 \BEA 
\mathbf{v}_i &=& v_0 \mathbf{e}_i \sigma_i (1 - s_i) + \xi \sum_{i \neq j} \mathbf{f}(r_{ij})\;,\label{eq:vi}\\
\w_i&=& \mathbf{t}^r_i \, s_i \, \sigma_i \;,\label{eq:omegai}
\EEA
where $\mathbf{v}_i=\partial_t\rr_i$ and $\w_i=  \mathbf{e}_i \times \partial_t\mathbf{e}_i$ 
are the translational and angular velocity of a particle at positions $\rr_i$. The orientation $\mathbf{e}_i$ specifies the  direction or motion during the run phase.
The first term on the right hand side of Eq.~\eqref{eq:vi} describes propulsion at speed $v_0$, with $s_i$ an auxiliary state variable 
that is $0$ during the run and $1$ in the tumble state. Details about the model can be found in \cite{paoluzzi2013effective,Paoluzzi14}.
\mcm{In} the tumbling state, particle $i$ receives a random torque $\mathbf{t}_i^r$ that
rotates the  direction of its orientation $\mathbf{e}_i$. Tumbles are Poisson-distributed with mean rate $\lambda$.
The second term on the right hand side of Eq.~\eqref{eq:vi} describes repulsive interactions, 
with  $\mathbf{f}(r_{ij})=-\nabla_i V(r_{ij})$ and $V(r)=\frac{\epsilon}{12} (\frac{a}{r})^{12}$ the interaction potential among two particles at separation $r_{ij}=|\mathbf{r}_i-\mathbf{r}_j|$, $\epsilon$ the energy scale and $\xi $ the mobility. We choose $\epsilon=1$ in all the following. 
For the case of  overdamped dynamics considered here, 
 particles exchange momentum instantaneously upon collision. 
Hence, collisions do not lead to 
 rearrangement of non-motile particles.

\section{Numerical results}\label{sec:numerics}

We have simulated the dynamics of $N$ particles in a
square box of side $L$, with periodic boundary conditions.
We have varied the 
 packing fraction $\phi=NA_s/L^2\equiv\rho A_s$, with $A_s=\pi  (a/2)^2$,  by varying the number of particles  $N$ at fixed $L$.
We  have  solved numerically Eqs.~\eqref{eq:vi},\eqref{eq:omegai}
using a second order Runge-Kutta scheme with time step $dt\!=\!10^{-3}$ and $\mu\!=\!v_0\!=\!1$. The 
results presented below
have been obtained considering $L\!=\!80 \,a$ and varying $\phi$ in the range $[1 \cdot 10^{-2} \div 0.79]$ for $\lambda\!=\!0.01,0.05,0.1,0.5,1,5$.
All simulations are initiated with one 
\marco{
\matteop{nonmotile ($R$)}
 particle in a sea of \matteop{motile ($G$)} particles
}
 \matteop{The logic interaction is turned on \mcm{after}   the system \mcm{has reached a} steady-state configuration \mcm{by evolving according to its active dynamics. This typically takes a simulation time larger than $10/\lambda$.}
}

\matteop{In Fig. (\ref{fig:psi02}) we \mcm{show} the phase diagram \mcm{obtained by}  numerical simulations for
two values of packing fraction. \mcm{A} generic feature \mcm{is the appearance of} two non-equilibrium phase transitions: a structural
phase transition due to MIPS, and a non-equilibrium absorbing state phase transition. 
In Appendix \ref{no-reawake} we report details about the case $\mu=0$, where
the system evolves towards an absorbing state whose color depends on the type of logic gate considered.}


\subsection{\matteop{Mixed state and MIPS}}

\matteop{When $\mu$ is small but finite,
non-motile particle\mcm{s}  {\it reawaken}, i.e., 
become motile again},  at an average rate $\mu$.
  
 We have studied numerically AND particles at two values of the total packing fraction: (i) high packing fraction  ($\phi=0.8$), where
one expects that  the relaxation towards the steady state may be captured by a mean-field description, and (ii)
intermediate packing fraction ($\phi=0.2$), where local density fluctuations become important.
We have constructed a phase diagram  by varying the tumbling rate $\lambda$ and the reawakening rate 
$\mu$, at fixed self-propulsion \mcm{speed $v_0$}.  Working at fixed density  allow\mcm{s} us to quantify the importance of
density fluctuations due to self-propulsion on the phase transition to the absorbing state.

\begin{figure}[!t]
\includegraphics[width=1.05\columnwidth]{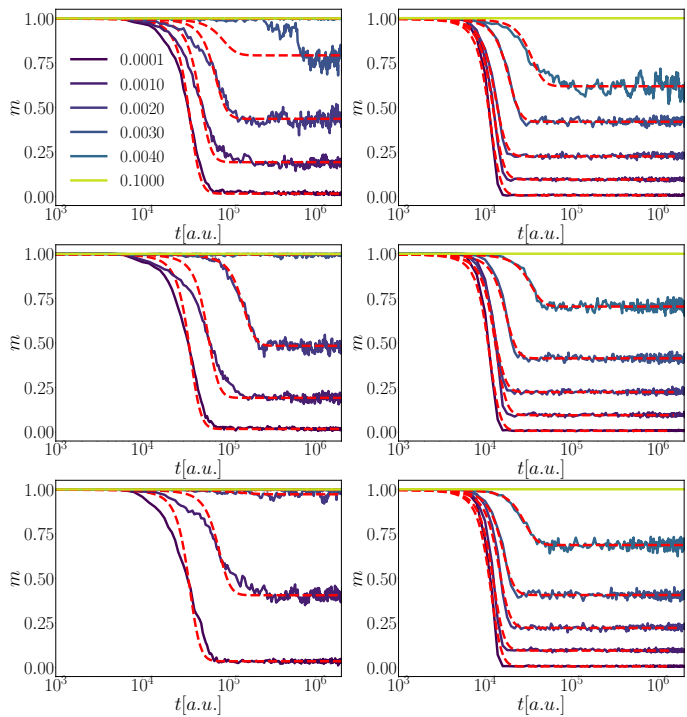}
\caption{ Time evolution of $m(t)$ as the reawakening frequency increases 
($\mu=10^{-4},10^{-3}, 2 \times 10^{-3}, 3 \times 10^{-3}, 4 \times 10^{-3}, 10^{-1}$, from violet to yellow) 
from green to violet, respectively), 
for $\phi=0.2$ (left column) 
and $\phi=0.8$ (right column) at different values of $\lambda$: $\lambda=0.01$ (top row), $\lambda=0.1$ (middle row) and $\lambda=1$ (bottom row).  Dashed red lines are fits to the mean-field model. 
\label{fig:mag_time02} 
}
\end{figure}

\begin{figure}[!t]
\includegraphics[width=.4935\columnwidth]{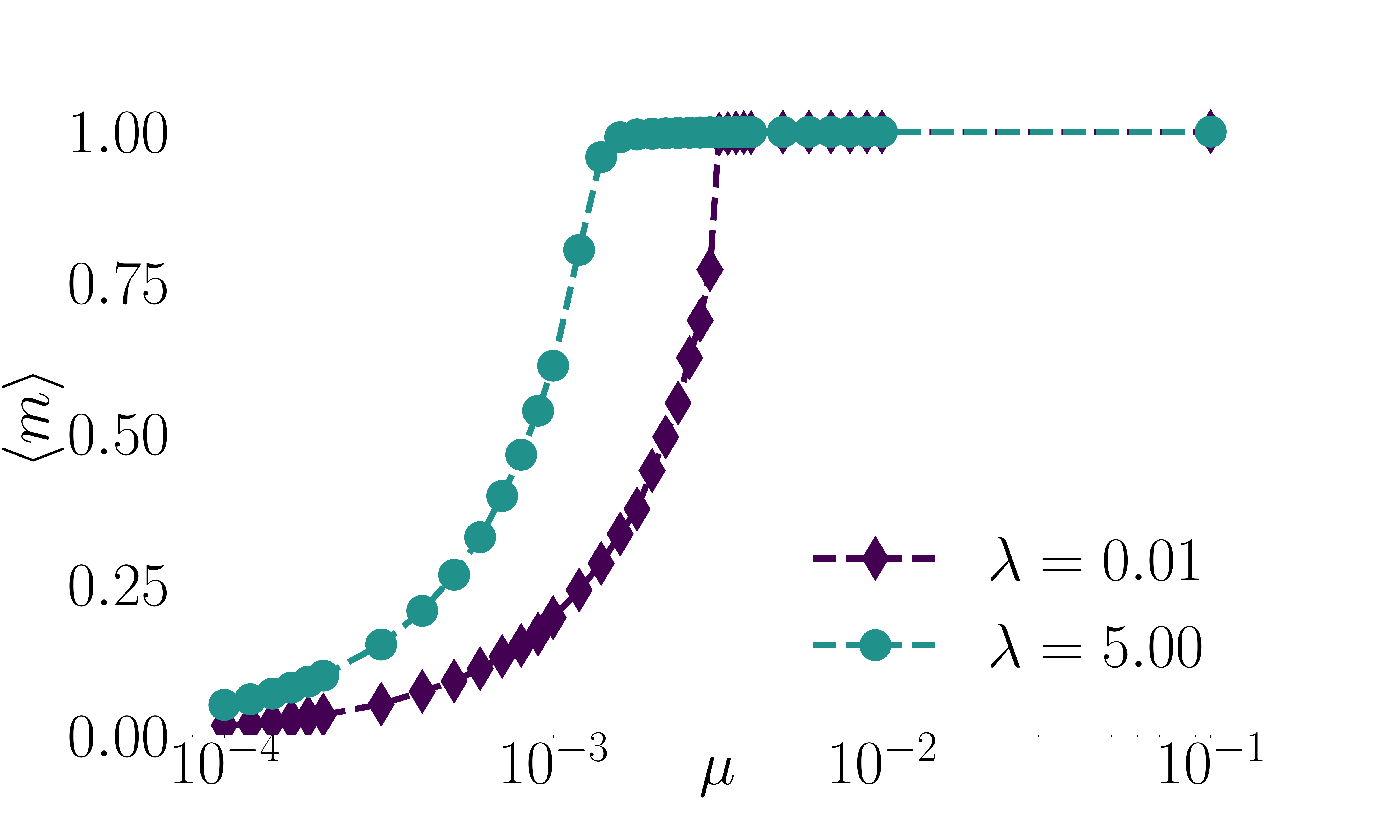} 
\includegraphics[width=.4935\columnwidth]{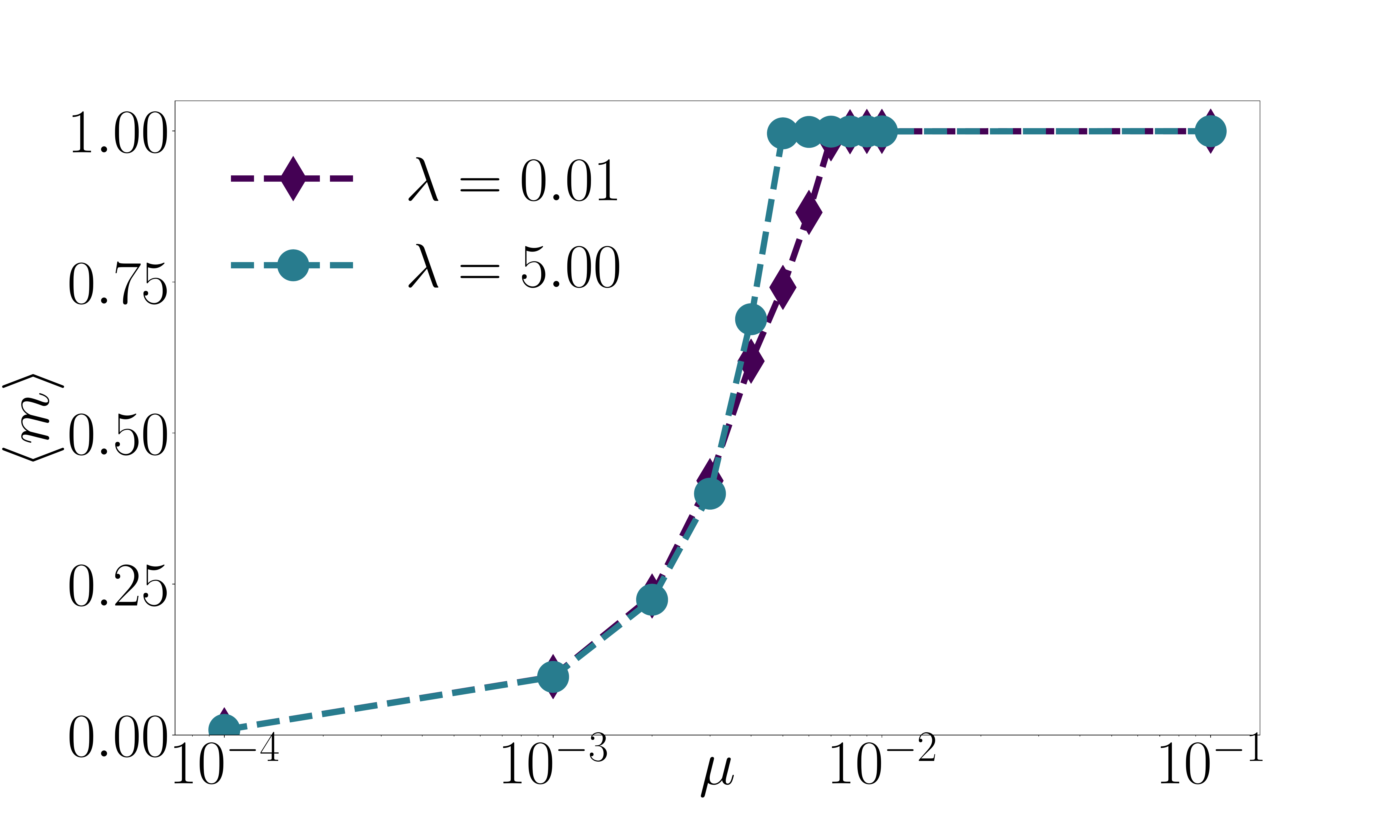} 
\caption{ 
The order parameter \matteop{$\langle m\rangle$}  as a function of $\mu$ for $\phi=0.2$ (left) and $\phi=0.8$ (right).}
\label{fig:mag} 
\end{figure}

Because of reawakening, the fraction of motile particles \matteop{$m(t)=N^{-1}\sum_i \sigma_i(t)$} fluctuates in time in the steady-state. 
For finite $\mu$,  the behavior is controlled by the interplay \mcm{between} reawakening and collisions. 
%
\mcm{S}tarting with one motile ($G$) particle in a sea on non-motile ($R$) particles, AND particles relax to one of two states: (i) an absorbing state
of all motile particles with 
\matteop{$\langle m \rangle =1$}  for $\mu>\mu_c$, referred to as  {\it motile state}, 
 and, (ii) a {\it mixed state} with
\matteop{$\langle m \rangle <1$}
 and a finite fraction  of non-motile particles for $\mu<\mu_c$. 
 \matteop{\mcm{Here } $\langle \mathcal{O}  \rangle$ \mcm{denotes} the time-average of the dynamical observable $\mathcal{O}(t)$
in the steady-state.}
 This behavior is evident in the relaxation of $m(t)$  shown in  Fig. (\ref{fig:mag_time02})  
 for $\phi=0.2$ (left) and $\phi=0.8$ (right).
The absorbing state of all non-motile particles (\matteop{$\langle m\rangle =0$}) is obtained  only for 
$\mu=0$. The structural and dynamical properties of this state were studied in Ref. \cite{Draft-Matteo}. 
The behavior of \matteop{$\langle m\rangle $} is shown in Fig. (\ref{fig:mag}) as a function of $\mu$ for the two representative packing fractions.

The dashed red lines  in Fig.~\ref{fig:mag_time02} are fits to the analytical mean-field solution 
\mcm{given in  Eq.~ (\ref{eq:mf_sol}) below, with }
\matteop{$\mu$ and $\tau$  fitting parameters}.
The mean-field model fits well at high packing fraction (right column, $\phi=0.8$), 
but fails at low packing fraction when density fluctuations are important (left column, $\phi=0.2$).
 %

\mcm{S}napshots \mcm{of the long-time configurations shown in Fig.~\ref{fig:snap} display} a clear propensity of particles to aggregate. The aggregation of motile green particles at high density is  a manifestation of motility-induced phase separation (MIPS) that is most pronounced at low tumbling rates, when the single-particle dynamics is persistent. At $\phi=0.2$ no MIPS is expected in the range of parameters considered here. 
Snapshots of the absorbing state at low reawakening for $\phi=0.2$ show a different type of aggregation, corresponding to  phase separation between green and
red particles, with   a compact cluster of non-motile red particles surrounded by a gas of motile green
particles.  

\begin{figure}[!t]
\includegraphics[width=.475\columnwidth]{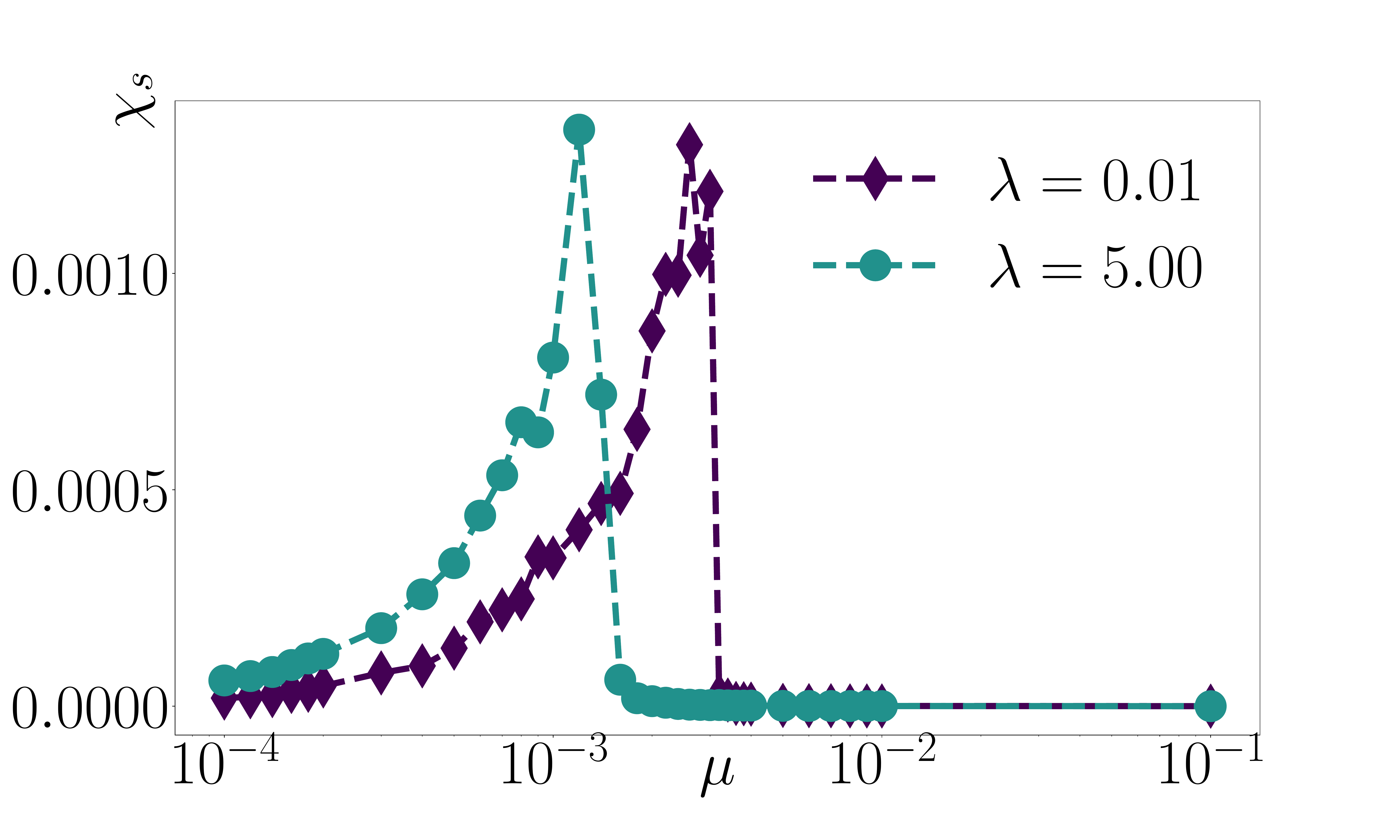} 
\includegraphics[width=.475\columnwidth]{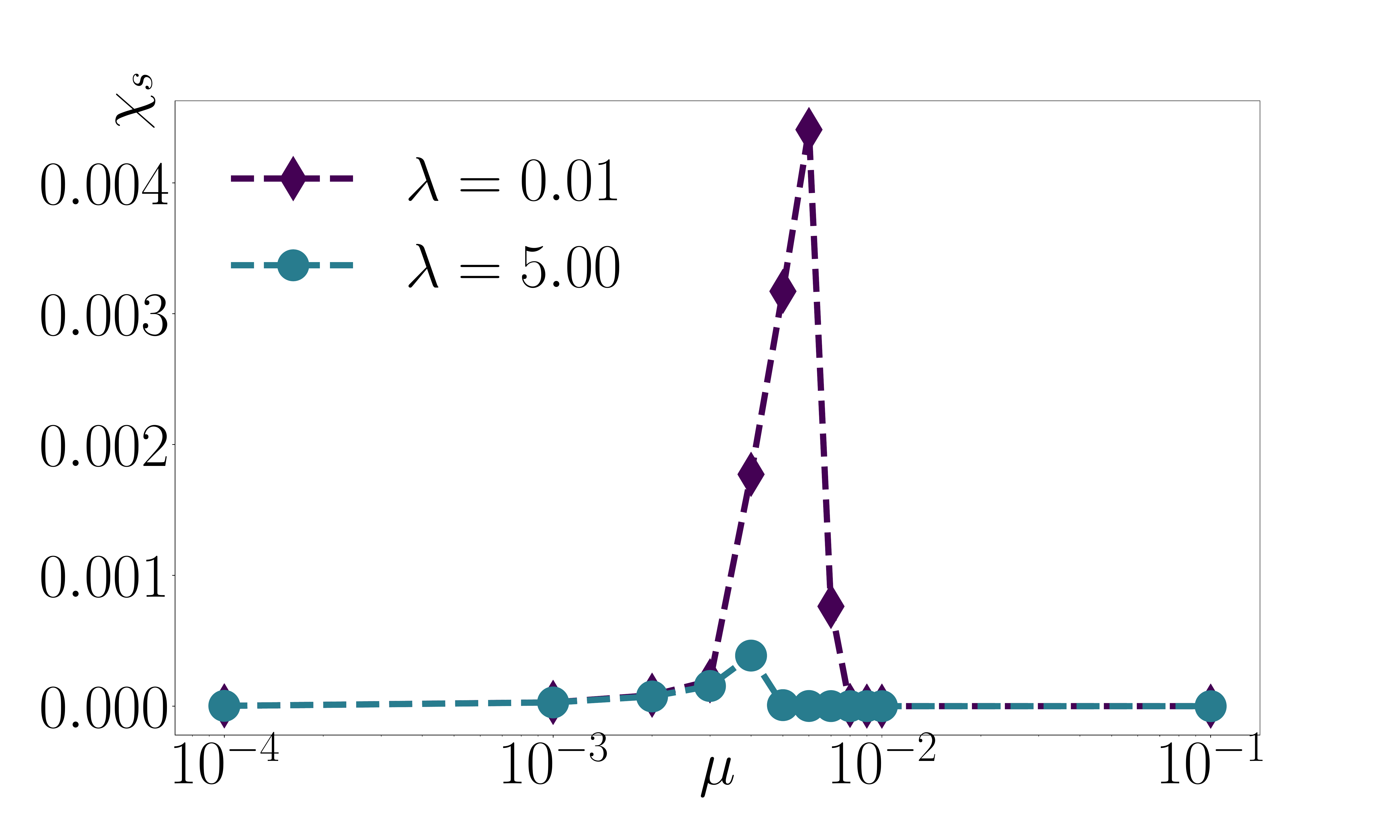} 
\caption{  
Steady-state susceptibility $\chi_s$ as a function of $\mu$ for $\phi=0.2$ (left) and $\phi=0.8$ (right).
The position of the peak depends on $\lambda$ for $\phi=0.2$, while  it is essentially independent of $\lambda$ for $\phi=0.8$.}
\label{fig:magchi} 
\end{figure}

\mcm{The phase diagrams shown in Fig.~\ref{fig:psi02} } for $\phi=0.2,0.8$ have been obtained for
 $\lambda \in [0.01,5]$ and $\mu  \in [10^{-4},10^{-1}]$.
To quantify the transition  between motile and mixed states, 
we have examined 
\matteop{$\chi_s = \lim_{t \to \infty} \chi(t)$ where we have introduced $\chi(t)\equiv \langle (  m_s(t) - \langle m \rangle_s )^2 \rangle_s$,
with $\langle\cdots\rangle_s$  denoting a sample average.
$\chi(t)$ is analogous to, but distinct from, the 4-point susceptibility used in glassy physics that measures  sample-to-sample fluctuations of the 
overlap between configurations accessed by the system at different times~\cite{Lacevic2003}. The typical behavior of $\chi(t)$ is shown in Appendix \ref{no-reawake}.}
\matteop{$\chi_s$ is the long-time limit value of $\chi(t)$, it does not provide any information about dynamics, however,
since it measures the fluctuations of $\langle m \rangle$ in the steady-state, it is suitable for identifying regions in the phase diagram where $\langle m \rangle$ changes 
from motile to mixed state. The behavior of $\chi_s$}
is shown in Fig.~(\ref{fig:magchi}) as a function of $\mu$ for different values of $\lambda$. 
We identify $\mu_c$ 
with the location of the peak in the susceptibility \mcm{(blue diamonds in Fig.~\ref{fig:psi02})}. 
At intermediate density the location of the peak shifts to lower values of $\mu$ with increasing tumbling rate, while at high density it is essentially independent of tumbling rate.


\begin{figure}[!t]
\includegraphics[width=.475\columnwidth]{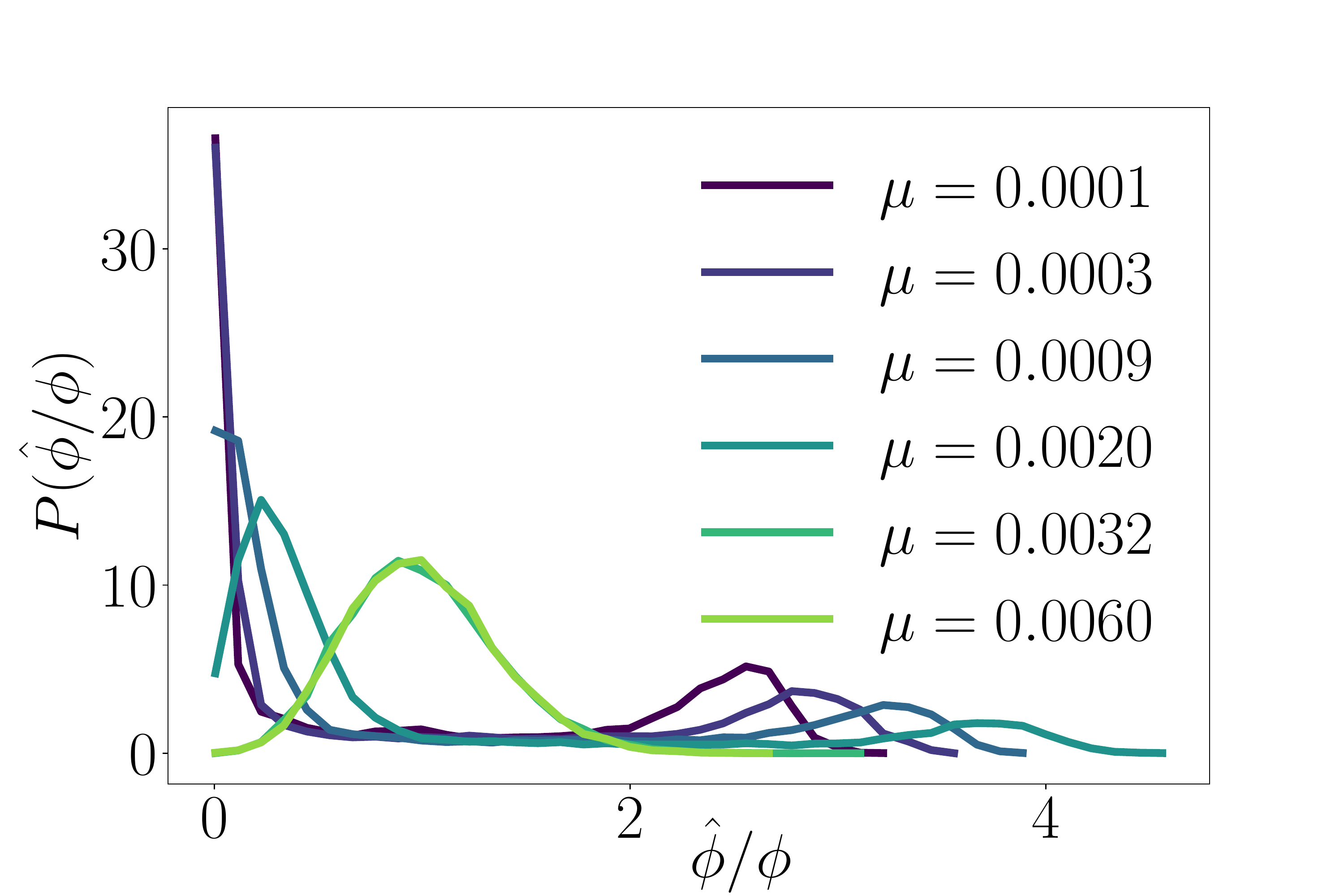}  
\includegraphics[width=.475\columnwidth]{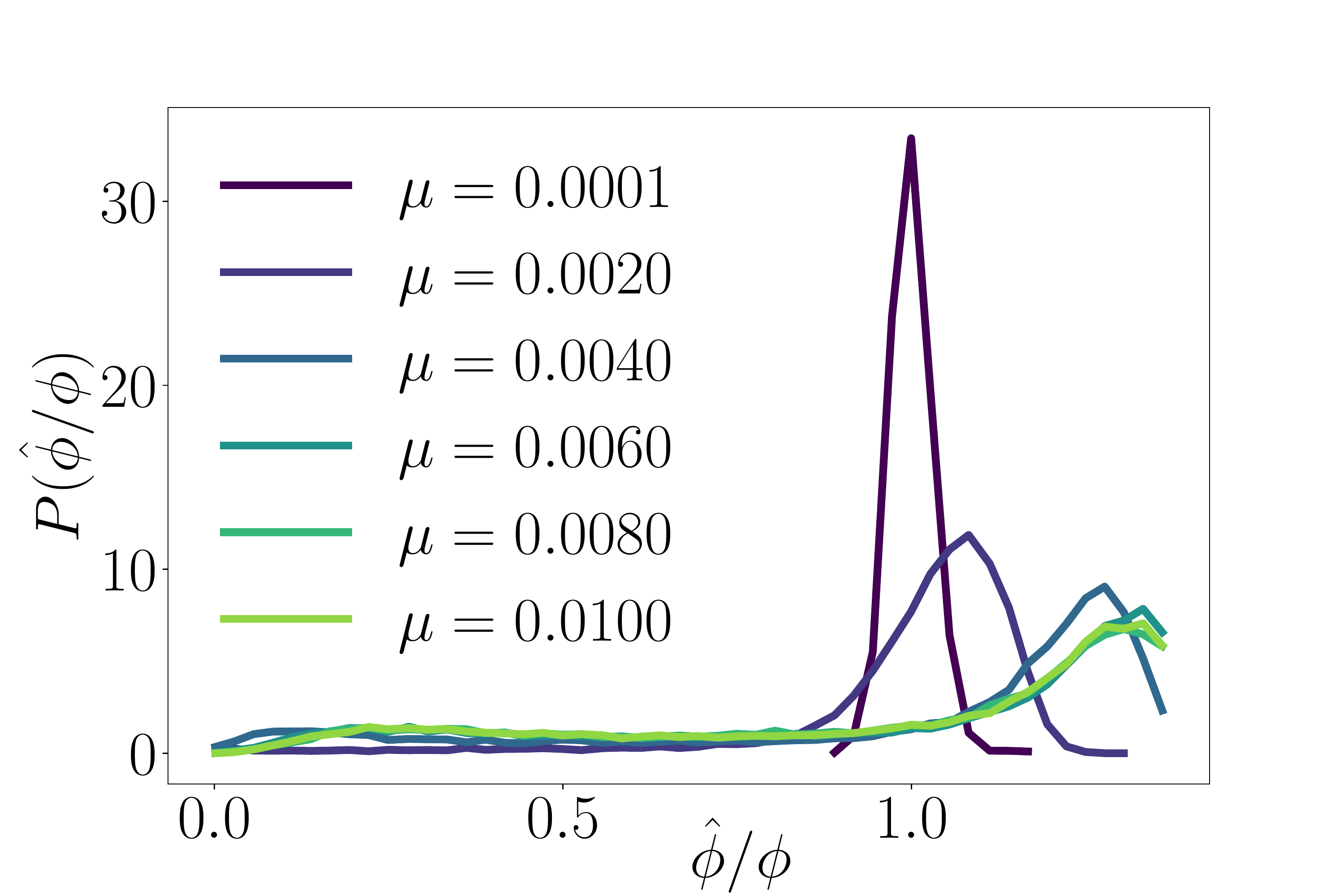}  
\caption{ 
Probability distribution function of density fields for $\phi=0.2$ (left) and $\phi=0.8$ (right) and $\lambda=0.01$.
\label{fig:pdr04} 
}
\end{figure}

To quantify the structural properties of the steady state, we have evaluated the probability distribution $P(\hat{\phi})$
of local  packing fractions $\hat{\phi}$ \matteop{(details are provided in Appendix \ref{pdf_pack}). }
shows an evolution from a unimodal distribution corresponding to a homogeneous state to a bimodal distributions corresponding to a phase separated state \mcm{(see  Fig. \ref{fig:pdr04}).} \mcm{T}he mechanisms driving \mcm{the} phase separation are, \mcm{however,} distinct at intermediate and high density. 
For $\phi=0.2$ one observes  phase separation between motile and non-motile particles in the asymptotic mixed state. This phase separation is driven by reaction-diffusion and weakly affected by $\mu$. For $\phi=0.8$ we find MIPS of motile particles in the absorbing state, hence the phase separation is driven by motility.
%
The black symbols \mcm{in Fig.~\ref{fig:psi02}}  have been obtained by calculating  the Binder cumulant $g_4$ of the 
distribution of local density $P( \hat{\phi} )$ defined as $g_4 = 1 - \langle \hat{\phi}^4 \rangle / 3 \langle \hat{\phi}^2 \rangle^2$ \cite{binder1981finite,rovere1988block,rovere1993computer}.
A negative peak in $g_4$  corresponds to a  bimodal $P(\hat{\phi})$, signaling a first-order phase transition \cite{binder1997applications}. 
%
 
 
At high densities\mcm{,} density fluctuations are suppressed and the transition line at $\mu=\mu_c$ is essentially independent of $\lambda$, confirming that the mean-field model 
provide\mcm{s} a good description of the system.
In contrast to what observed for $\phi=0.2$, the absorbing
state is highly homogeneous, as evident from the snapshots in Fig. (\ref{fig:snap}).
\matteop{For $\mu > \mu_c$, }
however, motile particles undergo  MIPS~\cite{Tailleur08,Fily2012,cates2015motility}. We note that at low $\lambda$, when 
\matteop{the active dynamics}
is most persistent, MIPS actually occurs even in the mixed phase, 
\mcm{provided}
the fraction of motile particles is sufficiently large.
 \matteop{This means that, for a fixed tumbling rate $\lambda$, MIPS disappears below a critical value \matteop{$\mu_c$}
and the system becomes homogeneous \mcm{upon} approaching the absorbing state.
 }

\section{Continuum Model} 
\mcmch{Here we formulate}  a continuum model \mcmch{of particles interacting with AND rules} that incorporates re-awakening. 
The continuum equations  \mcmch{ for the number densities $\rho_G$ and $\rho_R$ of motile (G) and non-motile (R) particles} are written as~\cite{Fily2014,Wittkowski2017}
\begin{align}
& \partial_t \rho_R     =  
\beta  \rho_G \rho_R - \mu \rho_R\; ,\label{eq:rhoR}\\
& \partial_t \rho_G        =   
- \beta \rho_R \rho_G + \mu \rho_R -\boldsymbol{\nabla} \cdot\left[v(\rho_G)\mathbf{J}\right]\;, \label{eq:rhoG}\\
& \partial_t \mathbf{J} (\rr,t) = -\lambda_{eff} \mathbf{J} -\frac{1}{2} \boldsymbol{\nabla}\left[ v(\rho_G)\rho_G\right] -\bm\nabla\kappa\nabla^2\rho_G\;, \label{eq:J}
\end{align}
where $\mathbf{J}$ is the current density of motile particles and $v(\rho_G)$ is \mcmch{their} propulsive speed
\mcm{that can be}  suppressed by crowding as in models of MIPS. We use the simple form
\begin{equation}
v(\rho_G) = v_0 \left(1 -\frac{\rho_G}{\rho^*}\right)\;, \quad \rho_G \le \rho^*
\label{eq:v}
\end{equation}
and $v(\rho_G)=0$ for $\rho_G>\rho^*$, where $\rho^*$ is a characteristic density that depends on particle motility, tumbling rate {\color{black} and strength of repulsive interaction}. It was estimated via kinetic arguments for instance in Ref.~\cite{Fily2014}. In our model 
non-motile particles are truly static and do not diffuse, and we neglect small cross-diffusion terms ~\cite{Wittkowski2017}. 
Collisions instantaneously change the state of a particle from motile to non-motile and are incorporated in the reaction  term proportional to $\beta$, hence do not contribute to the suppression of the propulsive speed. On the other hand, the reaction  \mcmch{kinetics that changes the particles' motility at rate} $\sim\beta$ renormalizes the tumbling rate of the motile agents to an effective tumbling rate $\lambda_{eff}=\lambda+\beta\rho_R$ (\matteop{see Appendix (\ref{sec:app-langevin}) for details}). 
\mcmch{This arises because collisions  with non-motile particles
  result in  an effective rotational diffusion rate $\sim\beta\rho_R$ of motile ones that adds to the tumbling rate.  }
We also neglect small cross-diffusion terms in the dynamics of motile particles. 
Finally, the last term on the right hand side of Eq.~(\ref{eq:J}) represents a phenomenological surface tension $\kappa>0$ that controls gradients in the density of motile particles.

On time scales long compared to $\lambda_{eff}$, we can neglect the left hand side of Eq.~(\ref{eq:J}) and eliminate the current from Eq.~(\ref{eq:rhoG}) to obtain an effective diffusion equation for the density of motile particles, given by
\BEQ
\partial_t \rho_G(\rr,t)        = 
 - \beta \rho_R \rho_G +\mu  \rho_R   
+\bm\nabla\cdot\left[\mathcal{D}(\rho_G)\bm\nabla\rho_G\right]-\kappa\nabla^4\rho_G\;,
   \label{eq:rhoGeff}
\EEQ
where 
\BEQ
\mathcal{D}(\rho_G)=\frac{v(\rho_G)}{2\lambda_{eff}(\rho_R)}\left[v(\rho_G)+\rho_Gv'(\rho_G)\right]\;,
\label{eq:Dgg}
\EEQ
with the prime denoting a derivative with respect to density. \mcmch{The  effective diffusivity  $\mathcal{D}$} incorporates the effects of crowding due to motility~\cite{Fily2012,Fily2014,cates2015motility}
and can change sign, signaling the spinodal instability associated with MIPS.  
  We therefore expect that at high enough density the structural properties of the system will be controlled by the interplay of MIPS physics and the exchange of internal motility regulated by collisions and reawakening.

\mcmch{\paragraph*{Mean-field model.} 
 Neglecting all spatial variations, Eqs. (\eq{eq:rhoR}) and \eq{eq:rhoG}) reduce to a logistic model augmented by re-awakening,
\BEA
\label{eq:MF}
\partial_t\rho_G&=&-\beta\rho_G\rho_R+\mu\rho_R\;,\\
\partial_t\rho_R&=&\beta\rho_G\rho_R -\mu\rho_R\;.
\EEA
The total density $\rho=\rho_G+\rho_R$ is constant, $\rho=\rho_0$, and the coupled equations can be recast in the form of a single equation for the fraction $m=\rho_G/\rho_0$ of motile particles, given by
\BEQ
\partial_tm=-\beta\rho_0 m (1-m)+\mu(1-m)\;.
\label{eq:phi}
\EEQ
The homogeneous steady states are controlled by the interplay between the time $\tau=(\beta\rho_0)^{-1}$ at which collisions turn motile particles into non-motile ones and the reawakening rate $\mu$.  One finds two stable states: an absorbing state of all motile particles ($m^*=1$) when \cris{$\mu>\mu_c=\tau^{-1}$ and a mixed state with a fraction $m^*=\mu/\mu_c$ of motile particles  for $\mu<\mu_c$.The rate $\mu_c=\beta\rho_0$ provides the mean-field value of the transition point between absorbing and mixed states.} In the absence of reawakening ($\mu=0$) the mixed state is an absorbing state where all particles are non-motile. We will see below that fluctuations not captured by the mean-field model  given in Eq.~(\ref{eq:phi}) can yield a rich structure for both the absorbing and mixed states.
The mean-field model can be solved exactly to obtain the kinetics of approach to the steady state, with the result
\BEQ \label{eq:mf_sol}
m(t)=\frac{m_0-\mu/\mu_c-\mu/ \mu_c (m_0-1)e^{(1-\mu / \mu_c) t \mu_c}}{m_0-\mu/ \mu_c-(m_0-1)e^{(1-\mu / \mu_c)t \mu_c }}
\EEQ
for a given initial fraction $m_0=m(t=0)$ of motile particles. Clearly $m(t)$ relaxes to $m^*=1$ for $\mu / \mu_c>1$ and to $m^*=\mu / \mu_c$ for $\mu /\mu_c$ for all initial values $m_0$, other than $m_0=1,\mu / \mu_c$. The kinetics of approach to the steady states is controlled by the shortest of the collision and reawakening times, and the dynamics can become very slow near the critical line separating the two steady states, where these two time scales are comparable.
}

\paragraph*{Stability of homoegeneous states.} We now
  examine the linear stability of the two homogeneous steady states to spatially varying fluctuations. We write $\rho_{G,R}=\rho_{G,R}^0+\delta \rho_{G,R}$, where  $\rho_{G,R}^0$ are the homogeneous fixed points, and expand to linear order in the fluctuations. Working in Fourier space, we let $\delta\rho_{G,R}=\sum_{\mathbf{q}}e^{i\mathbf{q}\cdot\rr}\hat{\rho}_{G,R}(\mathbf{q})$. 
 The linearized equations for the Fourier amplitudes are then given by
\begin{align}
& \partial_t \hat\rho_R     =  \left( \beta  \rho_G^0  - \mu
\right)\hat\rho_R+\beta \rho_R^0\hat\rho_G\;  
,\label{eq:drhoR}\\
& \partial_t \hat\rho_G        = \left(\mu-\beta\rho_G^0\right) \hat\rho_R  - \left(\beta  \rho_R^0  +\mathcal{D}(\rho_G^0)q^2+\kappa q^4\right)\hat\rho_G\;.
 \label{eq:drhoG}
\end{align}

\noindent\emph{Stability of motile state. } 
 When reawakening is faster than collisions ($\mu>\mu_c=\beta\rho_0$), the homogeneous state is the absorbing state  where all particles are motile, i.e.,   $\rho_G^0=\rho_0$ and $\rho_R^0=0$. 
 In this case 
 Eqs.~(\ref{eq:drhoR}) and (\ref{eq:drhoG}) are 
 decoupled. Letting $\hat\rho_{G,R}(\mcmch{\mathbf{q},}t)\sim e^{i\omega t} \hat\rho_{G,R}\mcmch{(\mathbf{q})}$, the dispersion relations of the relaxation rates are
\begin{align}
&i\omega_R=-\mu+ \marco{\mu_c} 
\;\\
&i\omega_G=-\left(\mathcal{D}\mcmch{(\rho_0)}q^2+\kappa q^4\right)\;,
\end{align}
where
\BEQ
\mathcal{D}(\rho_0)=    \frac{v^2_0}{2  \lambda   }\left(1- \frac{\rho_{0}}{\rho^*} \right)
 \left( 1   - 2   \frac{\rho_{0}}{\rho^*}  \right) 
\EEQ
becomes negative at $\rho_0=\rho^*/2$.
Fluctuations in the density of non-motile particles always decay. On the other hand, fluctuations in the density of motile particles grow when $\mathcal{D}(\rho_0)<0$. The motile particle\matteop{s} aggregate and undergo MIPS for
 $ \rho_0>\rho^*/2  $.  
 Therefore if $\rho^*<\mu/\beta$ the motile state  will be homogeneous for $\rho_0<\rho^*/2$ and will undergo MIPS for $\rho^*/2<\rho_0<\mu/\beta$.

\noindent\emph{Stability of mixed state. } When collisions dominate over reawakening ($\mu<\mu_c=\beta\rho_0$),  the final state always contains a fraction of non-motile particles, with $\rho_G^0=\mu/\beta$ and $\rho_R^0=\rho_0-\mu/\beta$. In this case the linearized equations become
\begin{align}
& \partial_t \hat\rho_R     =  \left(\marco{\mu_c}
-\mu\right)\hat\rho_G\; ,\label{eq:drhoR}\\
& \partial_t \hat\rho_G        =  - \left(\marco{\mu_c}
-\mu  +\mathcal{D}(\rho_G^0)q^2+\kappa q^4\right)\hat\rho_G\;.
 \label{eq:drhoG}
\end{align}
Fluctuations in the density of non-motile particles are slaved to those in the density of motiles ones, whose decay is controlled by the rate
\BEQ
i\omega_m=- \left(\marco{\mu_c}
-\mu  +\mathcal{D}(\rho_G^0)q^2+\kappa q^4\right)\;,
\label{omegam}
\EEQ
with
  \begin{equation}
  \mathcal{D}(\rho_G^0)=    \frac{v^2_0}{2 \left( \lambda+\mu\right) }
  \left(1- \frac{\mu }{\beta \rho^*} \right)
 \left( 1   - 2   \frac{\mu }{\beta \rho^*}  \right) \;.
 \label{eq:DrhoG}
  \end{equation}
The mixed homogeneous state is then unstable if the following conditions are satisfied
\BEQ
\mathcal{D}(\rho_G^0)<0~~~~~~~{\rm and}~~~~~~~|\mathcal{D}(\rho_G^0)|>\sqrt{4\kappa(\marco{\mu_c}
-\mu)}\;.
\label{eq:boundary}
\EEQ
The first condition is satisfied for 
$ \mu/\beta > \rho^*/2 $, 
 which provides a necessary, but not sufficient condition for the instability. 
%
%
\textcolor{black}{ At the onset of instability only one mode is unstable, with wavevector  $q_0=\left(|\mathcal{D}(\rho_G^0)|/2\kappa\right)^{1/2}$. Using Eq. (22), we can also write 
\begin{equation} \label{eq23}
q_0 = \sqrt {\frac{ \mu_c - \mu  }{ | \mathcal{D}(\rho_G^0)| }} {\sim(\mu_c-\mu)^{1/2}}\;.  
\end{equation}
The wavevector $q_0$ sets the length scale of the spatial pattern. 
Combining, Eq. (23) and Eq. (21) one finds that \mcm{after the onset of MIPS ($\mu>\beta \rho_* /2$) }
  $q_0$  decreases monotonically \mcm{as $\mu\rightarrow\mu_c^-$, and the length scale of the resulting pattern grows with increasing $\mu$, diverging at $\mu_c$.}}

\begin{figure}[h!]
\centering
\subfigure{ \includegraphics[scale=0.38]{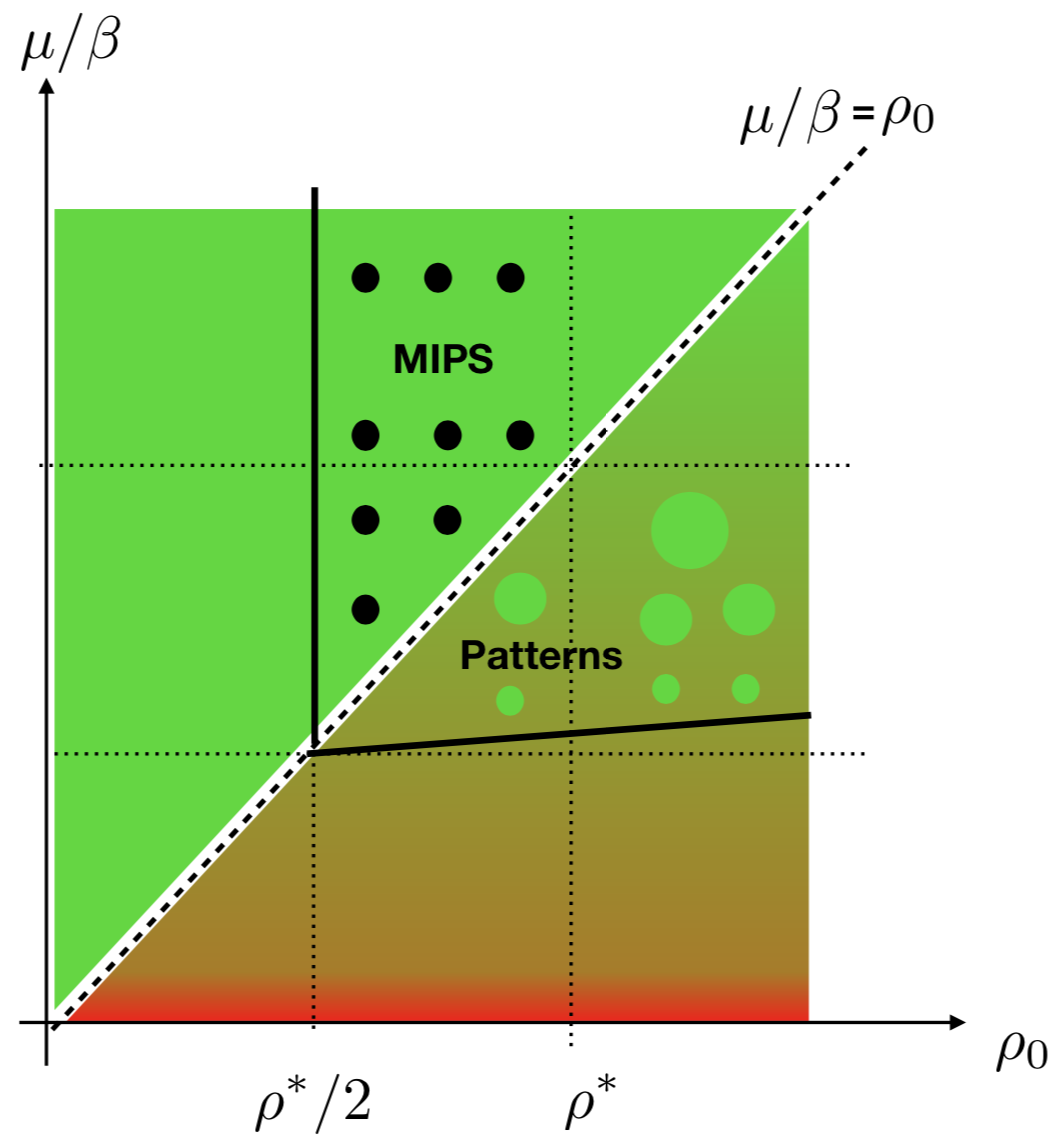} }
\subfigure{ \includegraphics[scale=0.55]{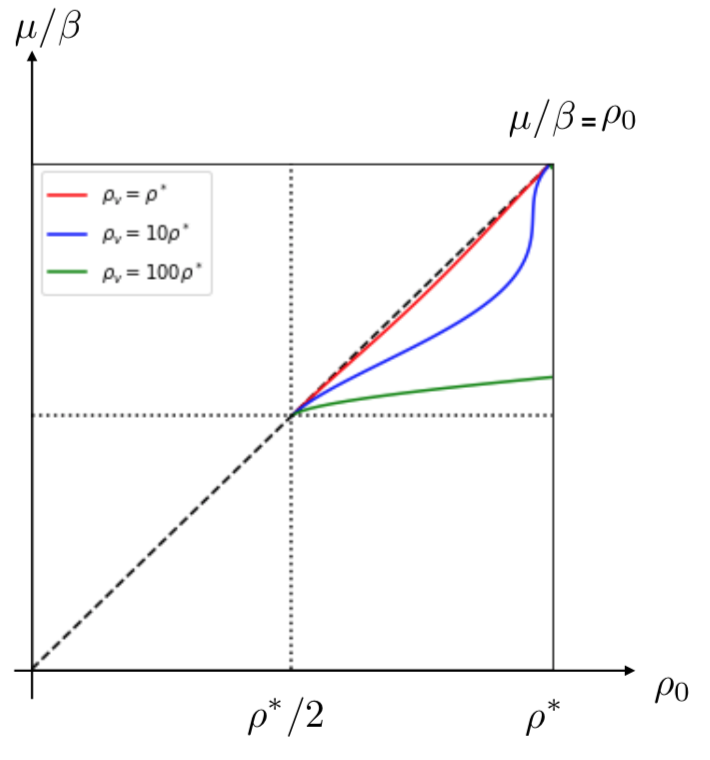} }
\caption{
Top: 
phase diagram obtained from linear stability analysis of the continuum model.
The  motile steady  state for $\mu/\beta>\rho_0$ is homogeneous  for 
$ \rho_0 <  \rho^*/2$, 
while the motile (G) particles undergo
  MIPS for $ \rho_0 > \rho^*/2$.
 In the mixed state for \cris{$\mu<\mu_c$} motile and non-motile particles are uniformly mixed for values of the reawakening frequency below the critical line determined by the solution of ~\eq{eq:boundary}
 and shown in the figure for $ \rho_v=v^4_0/(16 \lambda^2 \kappa \beta) $.
Above this line, fluctuations in the density of both motile and nonmotile particles are unstable, resulting in spatial patterns.
    Bottom: stability boundaries 
 in the mixed state for different values of the characteristic density 
    $\rho_v=  \rho^*, 10 \rho^*, 100 \rho^*$ ( with $\mu /\beta $,  and $\rho_0$ also measured in units of $\rho^*$).
 For simplicity we have neglected  the renormalization of $\lambda$. 
    } \label{fig:model}
 \end{figure}

In this regime fluctuations in the density of non-motile particles are slaved to those in the density of motile ones. As a result, both go unstable for $\rho_0>\rho^*/2$ above a critical value
\marco{$\mu_m(\rho_0)$} of the reawakening rate given by the solution of Eqs.~(\ref{eq:boundary}). As a result, the system shows regions that are essentially void of particles in a well mixed background (Fig. \ref{fig:model}). 
This is seen in our numerical simulations at high density. 
Although the numerics do not seem to show the emergence of patterns at a characteristic length scale, the size of the voids does increase with increasing $\mu$ 
as seen in the \mcmch{fourth and fifth}  column of Fig.~\ref{fig:snap} and consistent with 
 \eq{eq23}.

\section{Conclusions}
We have used numerics and mean-field theory to study a model  of active particles carrying  a Boolean variable coupled to their motility. 
When the motility state is exchanged irreversibly according to AND and OR logic rules, the system always evolves towards an absorbing state where all particles are 
motile (OR) or non-motile (AND), as shown in earlier work~\cite{Draft-Matteo}.
The coupling between motility and the spreading of information affects this irreversible dynamics, as evidenced by the asymmetry of the relaxation.
OR particles relax faster than AND particles, as shown in Fig.~\ref{fig:relax}, because their collisional  reaction rate is enhanced by activity, consistent with a  simple mean-field estimate. 
The key role of motility is also highlighted by examining a model where the internal state that evolves according to Boolean rules is simply the particles' color, while the particles remain motile at all times. 
In this case the relaxation of AND and OR particles is identical.

When AND particles are allowed to reacquire their motility at a rate $\mu$, we find both absorbing and active steady states controlled by the interplay of collision and reawakening rates.
For $\mu$ above a critical value $\mu_c$ the system evolves towards an absorbing state where all particles are motile. For $0<\mu<\mu_c$ the system evolves towards an active state with finite fractions of motile and non-motile particles, recovering the absorbing state of non-motile particles only at $\mu=0$. The value of $\mu_c$ is controlled by the total packing fraction and the collision rate. 
The steady state exhibits a rich spatial structure, with motility-induced phase separation \cris{(MIPS)} of motile particles in the high density motile absorbing state and aggregation of non-motile particles or void formation in the active mixed state.
Some of this behavior is reproduced by a mean-field model that incorporates suppression of motility due to crowding as in models of MIPS.

We have focused our attention on pattern formation in the mixed state. 
We showed that different types of aggregates \mcm{are} developed by logic interaction\cris{s}. 
At intermediate densities, the formation of aggregates is driven by the reaction-diffusion process that 
leads to the spreading of a cluster of nonmotile particles. At high densities, the phase separation is
driven by motility. In both cases, the non-equilibrium structural phase transition between homogeneous 
and phase-separated state\mcm{s} shows features of a first-order phase transition that is signaled by a negative peak of the Binder cumulant $g_4$. 
We leave the detailed study of the absorbing state phase transition for  future work. 

The simple model studied here provides a step towards understanding the role of motility in  information spreading. Specifically, the model of AND particles with reawakening is analogous to SIS models of epidemic spreading, but, in contrast to most existing studies, where  infection is spread on a static network, here we examine the case where the infection is spread by \emph{motile} agent\cris{s} and demonstrate that motility affects both the dynamics and the structure of the final state. The interplay of information spreading and motility is also relevant to pattern formation in bacterial colonies containing phenotypes with different motility, such as single- and multi-flagellated {\it Pseudomonas aeruginosa}\cite{Deforet} or  {\it B. Subtilis}, where a crossover from 
fractal  to compact bacterial aggregates has been observed \cite{Matsushita}, as well as to \mcm{the} understanding of biofilm formation.

\subsection*{Acknowledgements}
{
MCM was supported by the US National Science Foundation through award DMR-1609208. MP acknowledges funding from Regione Lazio, Grant Prot. n. 
85-2017-15257 ("Progetti di Gruppi di Ricerca - Legge 13/2008 - art. 4").
This work was also supported by the Joint Laboratory on
``Advanced and Innovative Materials'', ADINMAT, WIS Sapienza (MP). 
}



\appendix


\matteop{
\section{No reawakening and absorbing states} \label{no-reawake} 


We first examine the results of the model with $\mu=0$. Clearly in this case for both AND and OR rules the system evolves irreversibly towards an absorbing state with  all red (AND) or all green (OR) particles. 
It is clear from the rules given in Table~\ref{table1} that if the particles were to only exchange their color upon collision, while retaining their motile state, the two sets of rules would be symmetric when interchanging green and red. In this case the relaxation of AND and OR particles will be identical. Exchanging motility removes this symmetry and affects the dynamics.
\begin{figure}[t!]
\centering
\includegraphics[scale=0.275]{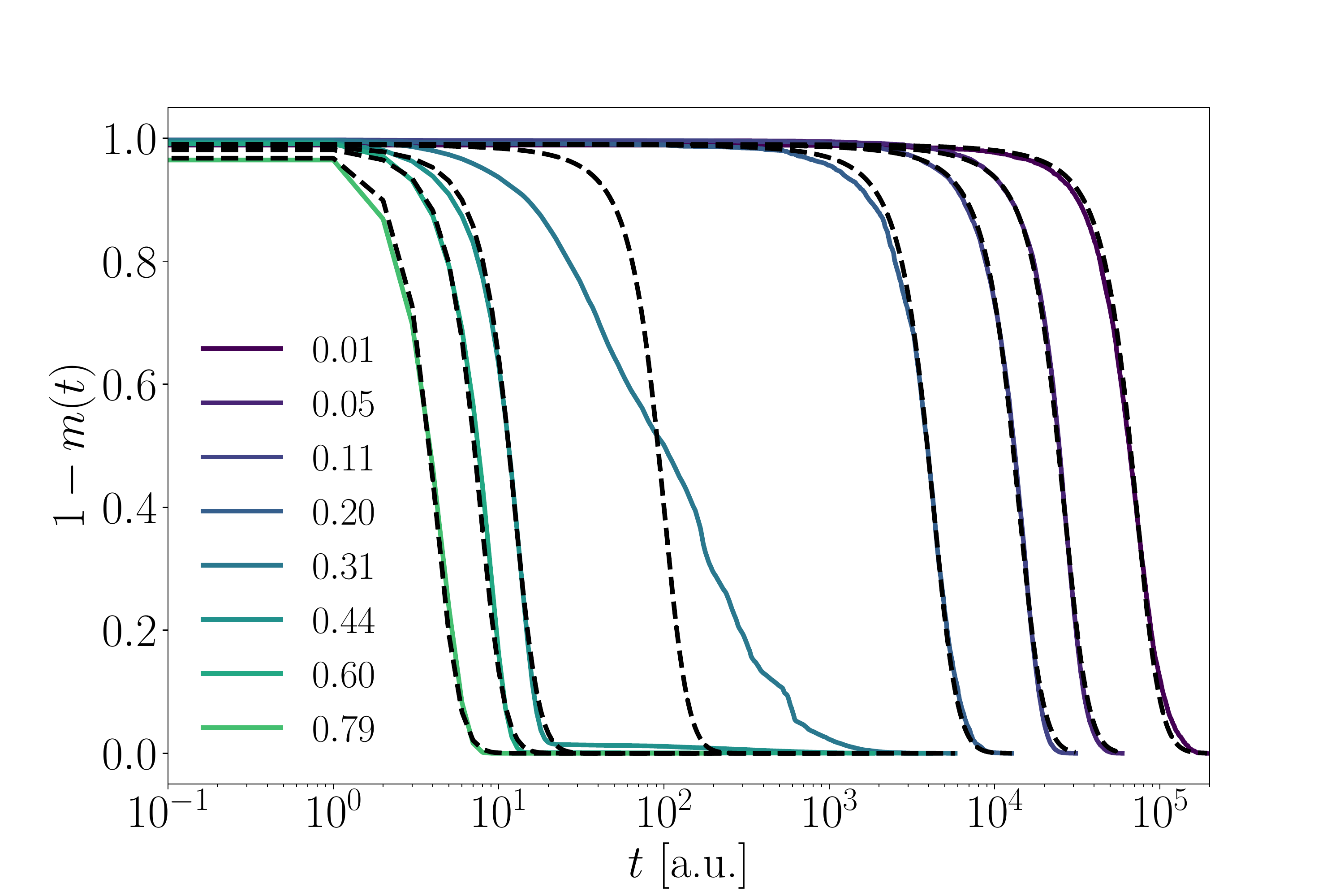} \\
\includegraphics[scale=0.275]{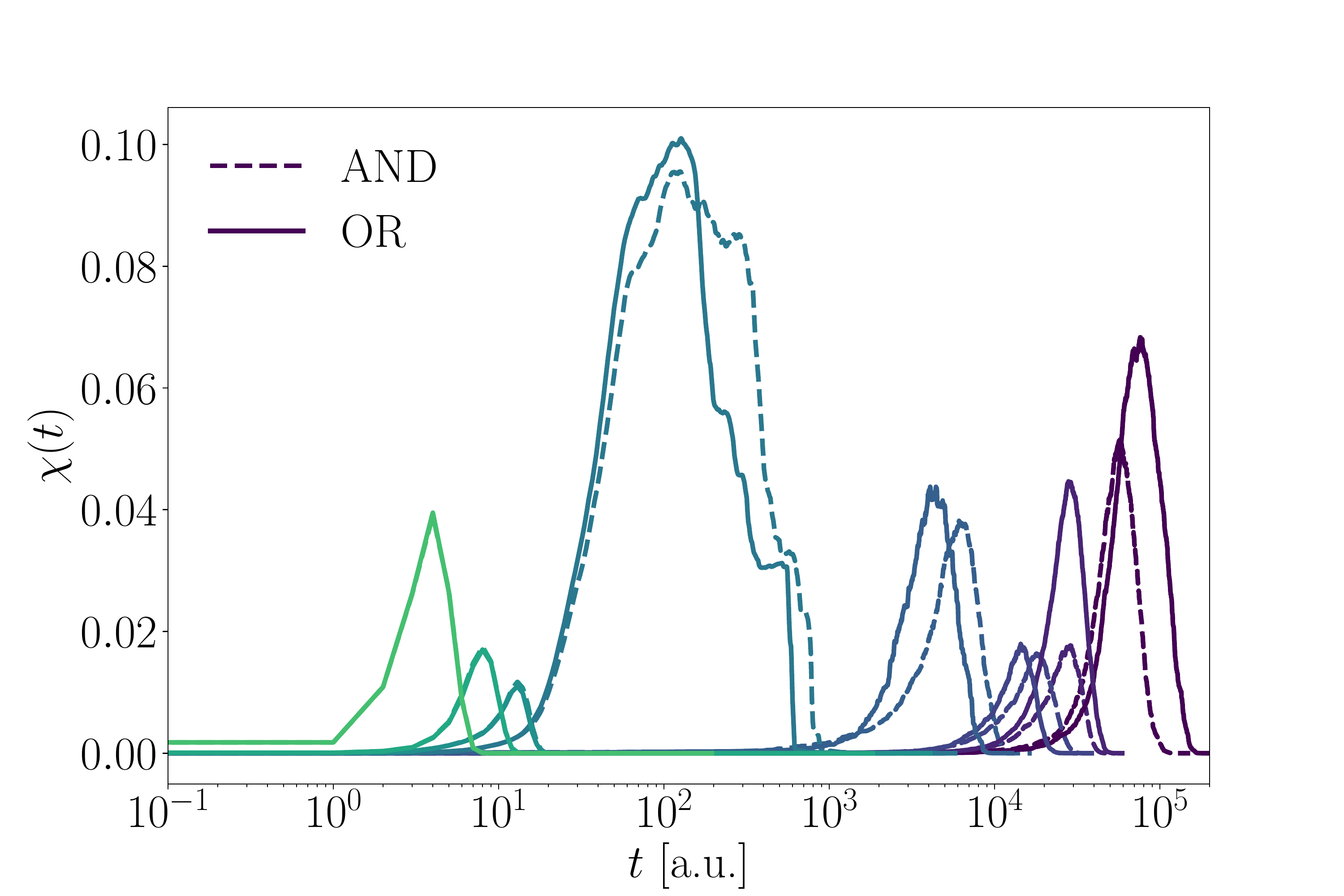}
\caption{ Top:   fraction $1-m(t)$ of motile OR particles  as a 
function of time  for various values of the packing fraction $\phi$.
The curves through the data are fits to the logistic form,  Eq. (\ref{eq:mf_sol}).
Bottom: Susceptibility
$ \chi(t)$ as a function of time (in arbitrary units) obtained from
 sample-to-sample fluctuations for AND (dashed) and OR (solid) particles. 
 The different curves are for the same values of packing fraction used in the top panel. Both figures are for $\lambda=1$ and $\mu=0$.}
  \label{fig:magnetization}
\end{figure}

To quantify the relaxation kinetics, we measure the fraction $m(t)$ of
 motile particles. 
The evolution of  $1-m(t)$ for OR particles  is shown in Fig.~\ref{fig:magnetization} (top panel) for different densities. At low and high density the dynamics is well reproduced 
by the logistic model given in Eq.~(\ref{eq:mf_sol}) with $\mu=0$ shown as dashed lines.
 The logistic model fails, however,  at intermediate densities. The evolution of $m(t)$ for AND particles was shown in Ref.~\cite{Draft-Matteo}.
\begin{figure}[t!]
\centering
\includegraphics[scale=0.275]{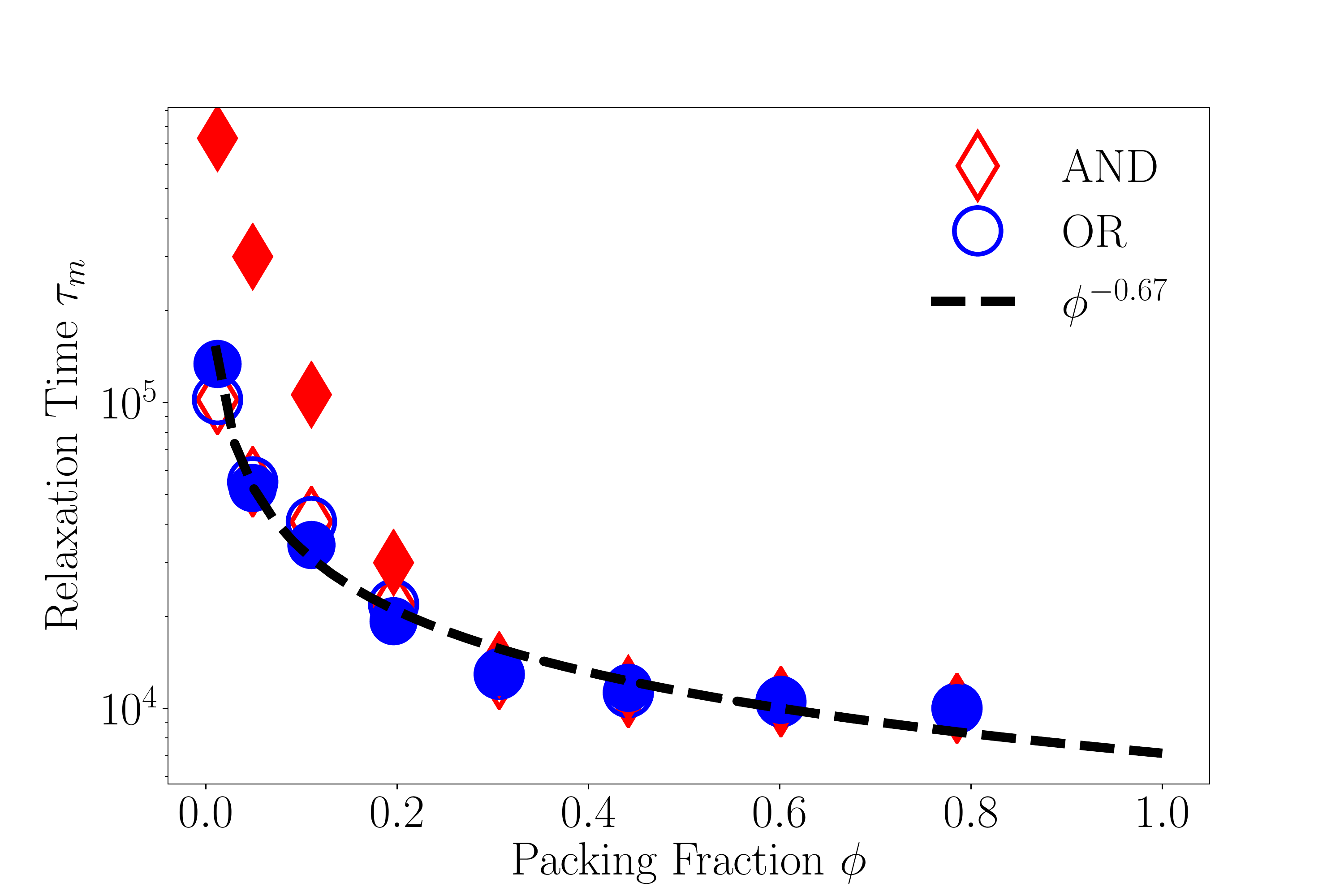}
\caption{ Relaxation time, $\tau_m$,  as a function of packing fraction. 
Red diamonds refer to AND particles, blue circles to OR particles.
Filled symbols correspond to the case where particles change their motility state upon collision according to the logic rules of Table~\ref{table1}. 
Open symbols are for the case where particles only exchange their color upon collisions, but always retain their motility. 
When the internal state and motility are decoupled 
the relaxation to the steady state  is the same for AND and OR rules.
When motility is exchanged according to the logic rules, OR particles relax faster than AND ones.
The black dashed line is a fit to  $\tau_m \sim \phi^{-0.67}$.
} \label{fig:relax}
\end{figure}
The fact that the logistic model does not  fit the data at intermediate densities indicates 
that in this regime the
relaxation dynamics cannot be described by a single time scale.
The existence of a distribution of relaxation times is highlighted by 
the dynamical susceptibility $\chi(t)$.
\cris{The height of $\chi(t)$ represents the variance of $m(t)$ at a given time, and thus the higher the peak the farther a given sample is from the average state. 
The growth in the height of the $\chi(t)$ peak seen in Fig. ~\ref{fig:magnetization} then reflects sample-to-sample fluctuations in the relaxation time. The non-monotonic behavior of the 
peak height with density arises because the distribution of relaxation times becomes narrow at both low and high density, where the mean-field logistic model  fits the data.}
%
We note that the complex kinetics obtained at intermediate density is not associated with the coupling between AND/OR rules and motility, 
but it also occurs for particles that only exchange color upon collision, while remaining always motile. 
The broad distribution of relaxation times arises instead 
 from the anomalous density fluctuations that are a signature of active systems.

Finally, we extract a mean relaxation time $\tau_m=\langle \hat\tau \rangle_s$, where $\hat\tau$ is defined by $m(\hat\tau)=0$ for AND particles and  $m(\hat\tau)=1$ for OR particles. 
The mean time $\tau_m$ 
is shown in Fig.~\ref{fig:relax} as a function of the total packing fraction. The plot clearly shows the faster relaxation of OR particles at low packing fraction. At high packing fraction, the curves are essentially on top of each other, and also agree with the relaxation for the case when motility and  particle color are decoupled and all particles  retain their motility upon collision.
The difference in the relaxation of 
  OR and AND particles when the internal state is coupled to motility
can be easily understood  by noting that for AND rules
the density of
 non-motile particles 
  grows at a rate $\sim\beta\rho_0$ controlled entirely by collisions with other particles. 
In contrast, for OR rules motile particles can additionally diffuse due to their motility, hence density variations over a region $\ell$ grow at a higher rate
$\sim\beta\rho_0 +\frac{v^2_0}{\lambda \ell^2}$. 
When interactions only result in color exchange, while particles remain always motile, then all particles
 diffuse and the relaxation rates become identical, as shown in Fig.~\ref{fig:relax}.

}

\section{
Renormalization of tumbling rate by interactions}
\label{sec:app-langevin}

In this Appendix we show that the exchange of the particle state through collisions yields a renormalization of   the tumbling rate. For simplicity, we carry out the calculation for the case of  Active Brownian Particles (ABP) instead of Run-and-Tumble Particles (RTP). For ABP interacting with AND logic rules, we show that the rotational noise $D_r$ is renormalized by interactions to the value $D_r^{eff}= D_r \matteop{+} \beta\rho_R$, where $\beta$ is the collision rate per unit density. 
Correspondingly, for RTP with AND interaction rule the effective tumbling rate is $\lambda_{eff}=\lambda +\beta\rho_R$.
We also provide an estimate of the parameter $\beta$.

We consider $N$ ABP
 at positions $\m x_n$   interacting with purely repulsive interactions with an internal state described by
 $s_n = \pm 1 $ ($s_n =  1$ motile, $s_n =  -1$ non-motile)  and orientation $\hh u_n$.
 The internal state variable $s_n$ is related to the  state variable $\sigma$ used in numerical simulations through $\sigma_n = \frac{s_n+1}{2}$. 
The dynamics is described by coupled Langevin-like equations, given by
\begin{align}
& \dot{\m x}_n = \frac{(1 +s_n )}{2} v_0 \hh u_n \;,
 \label{eq:xdot}   
 \\
&  \dot{s}_n =   -\beta \sum_{m\neq n}  (s_n + 1) \delta(s_n + s_m)
 \delta(\m x_n -\m x_m)\;, \label{eq:sigmadot} \\
& \dot{\hh u}_n = 
\frac{(1 + s_n )}{2}  \hh z \times \hh u_n \sqrt{2D_r} \eta_n(t)\;.
 \label{eq:thetadot}  
\end{align}
 The  right hand side of  \eq{eq:xdot} describes self-propulsion, which is zero if $s_n = -1$ (non-motile). 
The effect of interactions is only included in   \eq{eq:sigmadot}, where interactions with non-motile agent  change the motility state of the particle. It is not included  in the translational dynamics  that has been considered elsewhere ~\cite{Wittkowski2017}. 
The factor $s_n +1$ in \eq{eq:sigmadot} accounts for the sign of the derivative, which must be
  negative when switching from $+1$ to $-1$.
    Finally,
 \eq{eq:thetadot} describes the dynamics of the orientation $\hh u_n$, with $D_r$ the rotational diffusivity and $\eta_n(t)$ Gaussian white noise with unit variance.

The parameter $\beta$ can be estimated as  
$\beta=(\rho_0 \tau_c)^{-1}$,
 where $\rho_0$ is the mean density of particles and $ \tau_c$  the mean free time between collisions.
For a system of particles  traveling with mean speed $\langle v \rangle$ and density $\rho_0$,
then $ \tau_c = (\rho \langle v \rangle \sigma_c )^{-1}$, where $\sigma_c$ is the collision cross section, determined by the form of the repulsive potential.
The mean-free path is then  $\ell_c =  \langle v \rangle \tau_c $. 

At high density, the mean free path is smaller than the persistence length, $\ell_p = v_0/D_r$, or  $\ell_c < \ell_p$, corresponding to  $\rho_0>D_r/(v_0\sigma)$, particles  travel ballistically between collisions,  hence $\langle v\rangle = v_0$. This gives $\beta\sim v_0\sigma_c\sim v_0\rho_0^{-1/2}$ where in the last approximate equality we have assumed that at high density $\sigma_c\sim\rho_0^{-1./2}$.

At low density, 
$\ell_c >  \ell_p$
and particles 
  travel diffusively with diffusion coefficient $D_0 = \frac{v^2_0}{2D_r} $.
This gives $\beta  \sim   v_0 \sigma_c$, independent of density.
In both limits our estimates are  consistent with the results presented in
  \cite{Draft-Matteo}. 

To derive continuum equations
 we focus on the dynamics of the one-particle probability density \cite{PhysRevLett.90.138102}, given by  
\begin{equation}
c(\rr, \hh u, s, t) = \langle  \delta ( \rr -\m x(t) ) \delta(   \hh u - \hh u(t) ) \delta( s - s(t) ) \ \rangle
\label{eq:conc}
\end{equation}
that measures the probability of finding a  particle with  position $\rr$, orientation $\hh u$ and internal state $s$ at time $t$.

The dynamics of the probability density is governed by  a Smoluchowski equation that can be derived by standard procedure from the 
Langevin equations  \eq{eq:xdot}  - \eq{eq:thetadot}, \cite{Zwanzig}
Due to binary collisions,
the equation for $c$ couples to the two particle distribution function 
$ c_2(\m x, \hh u, s, t; \m x', \hh u', s', t)$. 
Using a Boltzmann-type of approximation  \cite{Zwanzig},
   we treat the two microscopic densities as uncorrelated and let
 $ c_2(\m x, \hh u, s, t; \m x', \hh u', s', t) \approx
 c(\m x, \hh u, s, t) c(\m x', \hh u', s', t)$,
 with the result
   \begin{align}
  & \partial_t c=  \nonumber
  -  \frac{1+s}{2}   \bm{\nabla}_x  ( v_0 \hh u c) - 
 \frac{(1+s)^2}{4} D_r 
  \mathcal{R} \cdot  \mathcal{R} c
 \nonumber \\
  &
  +\partial_s\left(  \beta(s+1) c(\rr, \hh u, s, t)  \int  d \hh u'   
     c(\rr, \hh u', -s, t)  \right )\;, 
  \end{align}
 where 
    $\mathcal{R} = \hh u \times \frac{\partial }{\partial  \hh u}$ is a rotation operator.

 We then write the total concentration as the sum of the concentrations of motile and non-motile particles
  $c_{G,R} = \sum_{s} \frac{(1 \pm s)}{2} c \approx \int_{s} \frac{(1 \pm s)}{2} c$
 and  introduce 
densities $\rho_{G,R} = \int d\hh  u \, c_{G,R}$ 
and the 
polar vector 
$\m J = v_0 \int d\hh u \, \hh u c_{G} $.
We carry out the integrals in $s$ 
using  Leibnitz rule 
for functions defined on compact spaces, where
  $\int ds f(s) \partial_s g(s) = - \int d s g(s)  \partial_s [ f(s)]  $.
  Thus 
   we obtain
   \begin{align}
  & \partial_t c_G =  
  - \bm{\nabla}_\rr    (   v_0 \hh u  c_G) -   D_r  \mathcal{R} \cdot  \mathcal{R} c_G 
 \label{eq:cdotp}
 \\
 & -  \beta   c_G(\rr, \hh u, t)   \rho_R(\rr,  t)\;, \nonumber \\
  & \partial_t c_R  = 
      \beta     c_G(\rr, \hh u, t)   \rho_R(\rr,  t)\;.
   \label{eq:cdotm}
  \end{align}

 Finally, integrating  \eq{eq:cdotp} and \eq{eq:cdotm} over $\hh u $, we obtain
   \eq{eq:cdotp} and
     \eq{eq:cdotm},   
with the result
   \begin{align}
  & \partial_t \rho_G =  
  - \bm{\nabla}_\rr       \m J 
 -    \beta   \rho_G   \rho_R\;,
    \label{eq:cdotp2}\\
   &  \partial_t \rho_R =    \beta     \rho_G   \rho_R\;,
   \label{eq:cdotm2}\\
      & \partial_t \m J =  
  -  \frac{1}{2} \bm{\nabla}_\rr    (   v_0^2 \rho_G) -  (  D_r +
  \beta      \rho_R  
   ) \m J \;.
    \label{eq:cdotp3}
  \end{align}

\matteop{
\section{Probability distribution of density fields} \label{pdf_pack}}
\matteop{The structural properties of the system have been investigated \cris{by}
looking at the probability distribution function $P(\hat{\phi})$ of local packing fraction $\hat{\phi}$ obtained
discretizing the simulation box into a lattice of linear size $\ell=4 a$.} 
\matteop{The transition lines have been obtained studying the behavior of the binder cumulant $g_4$ as a function of
the control parameters $\lambda$ and $\mu$. $g_4$ is a continuous function of the control parameter\cris{s} 
along a second-order phase transition, while a negative jump \cris{in this quantity} typically signals a first-order phase 
transition. 
}


\bibliography{biblio-info-active} 
\bibliographystyle{rsc} 

\end{document}